\documentclass[conferemce]{IEEEtran}
\usepackage{graphicx}
\usepackage{epstopdf}
\usepackage{amsthm}
\theoremstyle{definition}
\usepackage{epsfig}
\usepackage{latexsym}
\usepackage{amsfonts}
\usepackage{here}
\usepackage{rawfonts}
\usepackage[latin1]{inputenc}
\usepackage[T1]{fontenc}
\usepackage{calc}
\usepackage{capitalgreekitalic}
\usepackage{url}
\usepackage{enumerate}
\usepackage{color}
\usepackage[tbtags]{amsmath}
\usepackage{amssymb}
\usepackage{upref}
\usepackage{epic,eepic}
\usepackage{times}
\usepackage{dsfont}
\usepackage{comment}
\usepackage{cite}
\usepackage{bbm}
\usepackage{bm}
\usepackage{amsmath}
\usepackage{makecell}

\allowdisplaybreaks[4]
\newtheorem{theorem}{\bf Theorem}

\usepackage{dsfont}
\newcounter{step}
\newlength{\totlinewidth}
\newenvironment{algorithm}{%
  \rule{\linewidth}{1pt}
  \begin{list}{}%
    {\usecounter{step}%
      \settowidth{\labelwidth}{\textbf{Step 2:}}%
      \setlength{\leftmargin}{\labelwidth}%
      \setlength{\topsep}{-2pt}%
      \addtolength{\leftmargin}{\labelsep}%
      \addtolength{\leftmargin}{2mm}%
      \setlength{\rightmargin}{2mm}%
      \setlength{\totlinewidth}{\linewidth}%
      \addtolength{\totlinewidth}{\leftmargin}%
      \addtolength{\totlinewidth}{\rightmargin}%
      \setlength{\parsep}{0mm}%
      \raggedright}}%
  {\end{list}%
  \rule{\linewidth}{1pt}}
\newcounter{substep}

  {\end{list}}

\newlength{\aligntop}
\newlength{\alignbot}
 \makeatletter
\renewenvironment{align}{%
  \vspace{\aligntop}
  \start@align\@ne\st@rredfalse\m@ne
}{%
  \math@cr \black@\totwidth@
  \egroup
  \ifingather@
    \restorealignstate@
    \egroup
    \nonumber
    \ifnum0=`{\fi\iffalse}\fi
  \else
    $$%
  \fi
  \ignorespacesafterend%
  \vspace{\alignbot}\par\noindent
} \makeatother

\IEEEoverridecommandlockouts

\usepackage{algorithm}%,algpseudocode}
\usepackage{algorithmic}
\makeatletter
\newcommand\semihuge{\@setfontsize\semihuge{19.3}{25}}
\makeatother

\makeatletter
\newcommand\semismall{\@setfontsize\semihuge{12.4}{15}}
\makeatother

\usepackage{subfigure}
\begin{document}
\title{Digital Over-the-Air Federated Learning in Multi-Antenna Systems}
\author{{Sihua Wang,} \emph{Student Member, IEEE}, {Mingzhe Chen,} \emph{Member, IEEE},
{Cong Shen}, \emph{Senior Member, IEEE}, 
{Changchuan Yin}, \emph{Senior Member, IEEE},
and {Christopher G. Brinton}, \emph{Senior Member, IEEE} \\
\thanks{\scriptsize Sihua Wang and Changchuan Yin are with the Beijing Laboratory of Advanced Information Network, and the Beijing Key Laboratory of Network System Architecture and Convergence, Beijing University of Posts and Telecommunications, Beijing 100876, China. Emails: \protect\url{sihuawang@bupt.edu.cn;} ccyin@ieee.org.}
\thanks{\scriptsize Mingzhe Chen is with the Department of Electrical and Computer Engineering and Institute for Data Science and Computing, University of Miami, Coral Gables, FL, 33146, USA. Email: mingzhe.chen@miami.edu.}
\thanks{\scriptsize Cong Shen is with the Charles L. Brown Department of Electrical and Computer Engineering, University of Virginia, Charlottesville, VA, USA, Email: \protect\url{cong@virginia.edu}.}
\thanks{\scriptsize Christopher G. Brinton is with the School of Electrical and Computer Engineering, Purdue University, West Lafayette, IN, USA, Email: \protect\url{cgb@purdue.edu}.}
\thanks{An abridged version of this paper appeared in the Proceedings of the 2023 IEEE Global Communications Conference (GLOBECOM).}
}
% make the title area
\maketitle

\pagestyle{headings} % 加入一个页眉

\thispagestyle{empty}

\begin{abstract}
In this paper, the performance optimization of federated learning (FL), when deployed over a realistic wireless multiple-input multiple-output (MIMO) communication system with digital modulation and over-the-air computation (AirComp) is studied. In particular, a MIMO system is considered in which edge devices transmit their local FL models (trained using their locally collected data) to a parameter server (PS) using beamforming to maximize the number of devices scheduled for transmission. The PS, acting as a central controller, generates a global FL model using the received local FL models and broadcasts it back to all devices. Due to the limited bandwidth in a wireless network, AirComp is adopted to enable efficient wireless data aggregation. However, fading of wireless channels can produce aggregate distortions in an AirComp-based FL scheme. To tackle this challenge, we propose a modified federated averaging (FedAvg) algorithm that combines digital modulation with AirComp to mitigate wireless fading while ensuring the communication efficiency. This is achieved by a joint transmit and receive beamforming design, which is formulated as an optimization problem to dynamically adjust the beamforming matrices based on current FL model parameters so as to minimize the transmitting error and ensure the FL performance. To achieve this goal, we first analytically characterize how the beamforming matrices affect the performance of the FedAvg in different iterations. Based on this relationship, an artificial neural network (ANN) is used to estimate the local FL models of all devices and adjust the beamforming matrices at the PS for future model transmission. The algorithmic advantages and improved performance of the proposed methodologies are demonstrated through extensive numerical experiments.

\end{abstract}


\IEEEpeerreviewmaketitle
\vspace{-0.1cm}
\section{Introduction}

Federated learning (FL) has been extensively studied as a distributed machine learning approach with data privacy \cite{DKK,CGHS00,SCV01,JYK02,DMPA02,GHSK00,MPAN}. During the FL training process, edge devices are required to train a local learning model using its collected data and transmit the trained learning model to a parameter server (PS) for global model aggregation. The PS, acting as a central center, can coordinate the process across edge devices and broadcast the global model to all devices. This procedure is repeated across several rounds until achieving an acceptable accuracy of the trained model.

Since the PS and edge devices must exchange their trained models iteratively over the wireless channels, FL performance can be significantly affected by imperfect and dynamic wireless transmission in both uplink and downlink. Compared to the PS broadcasting FL models to edge devices, edge devices uploading local models to the PS is more challenging due to their limited transmit power \cite{XCJH04,WRCX04,HXR04,MSH04,TBSL03}. To tackle this challenge, over-the-air computation (also known as AirComp) techniques have recently been integrated into the implementation of FL \cite{YJJD03,NM03,SC03,YKTM04,CHX04}. Instead of decoding the individual local models of each device and then aggregating, AirComp allows edge devices to transmit their model parameters simultaneously over the same radio resources and decode the average model (global model) directly at the PS \cite{YDS05,ZYYW05,WZY05}. However, most of these existing works, such as  \cite{XHQQ01} and \cite{HAV}, focused on the use of AirComp for analog modulation due to its simplicity for FL convergence analysis, which may not be desirable for practical wireless communication systems that almost exclusively use digital modulations. In consequence, it is necessary to study the implementation of AirComp-based FL over digital modulation-based wireless systems.

\subsection{Related Works}

Recent works such as \cite{MDS1,ZJY2,CSZ3,YMMZ4,SJY5,MTDS6,HXY4,YGC,BJCYZ,XGJK,FJVS} have studied several important problems related to the implementation of AirComp-based FL over wireless networks. The authors in \cite{MDS1} minimized the mean-squared error (MSE) of the FL model during AirComp transmission under transmit power constraints in a multiuser multiple-input multiple-output (MIMO) system. In \cite{ZJY2}, the authors maximized the number of devices that can participate in FL training under certain MSE requirements in an AirComp-based MIMO framework. A joint machine learning rate and receiver beamforming matrix optimization method was proposed in \cite{CSZ3} to reduce the aggregate distortion and satisfy an FL performance requirement. The authors in \cite{YMMZ4} investigated the deployment of FL over an AirComp-based wireless network to minimize the energy consumption of edge devices. In \cite{SJY5}, the authors optimized the set of participating devices in an AirComp-assisted FL framework to speed up FL convergence. A receive beamforming scheme was designed in \cite{MTDS6} to optimize FL performance without knowing channel state information (CSI). The authors in \cite{HXY4} minimized the FL model aggregation error under a channel alignment constraint in a MIMO system. The authors in \cite{YGC} derived the optimal threshold-based regularized channel inversion power control solution to minimize the mean squared error (MSE) for an analog AirComp system with imperfect CSI. A channel inversion-based power control policy was developed in \cite{BJCYZ} to resist channel fading and meet the SNR threshold in an AirComp FL framework. In \cite{XGJK}, the authors minimized the computation error by jointly optimizing the transmit power at devices and a signal scaling factor at the PS. The authors in \cite{FJVS} minimized the transmit power of each device while ensuring a minimum MSE performance. One key challenge faced by these works is their limited practical applicability to high-order digital modulation-based wireless systems. By harnessing high-order digital modulation, individual devices can encode FL model parameters into discrete symbols and transmit multiple symbols simultaneously within the same frequency band. This approach optimizes bandwidth utilization and enhances data transmission efficiency. Moreover, digital modulation leverages error correction coding and modulation schemes (such as QAM) to enhance resistance against noise and interference. Hence, it is important to note that these works do not explicitly consider the incorporation of coding and digital modulation techniques, which are crucial components in real-world wireless communication systems.

Recently, several works \cite{GYDK6,RS7,XLR8,MD9,ZXVY9,XSB9,ABS,SJC,LCDHX} have studied the implementation of AirComp FL over digital modulation-based wireless systems. The authors in \cite{GYDK6} designed one-bit quantization and modulation schemes for edge devices. One-bit gradient quantization scheme is proposed in \cite{RS7} to achieve fast FL model aggregation. In \cite{XLR8}, the authors designed a joint channel decoding and aggregation decoding scheme based on binary phase shift keying (BPSK) modulation for AirComp FL. The authors in \cite{MD9} evaluated the performance of FL gradient quantization in digital AirComp. In \cite{ZXVY9}, the convergence of FL implemented over an AirComp-based MIMO system is derived. The authors in \cite{XSB9} proposed a digital transmission protocol tailored to FL over wireless device-to-device networks. In \cite{ABS}, the authors proposed an AirComp method that utilizes non-coherent detection and digital modulation to achieve high test accuracy. The authors in \cite{SJC} adopted BPSK for over-the-air computations to minimize the normalized MSE. A joint transmission and local computing strategy was designed in \cite{LCDHX} that utilizes multiple amplitude shift keying (MASK) symbols to minimize the energy consumption used for FL training. However, these prior works \cite{GYDK6,RS7,XLR8,MD9,ZXVY9,XSB9,ABS,SJC,LCDHX} mainly used low order digital modulation (i.e., BPSK) and hence their designed AirComp FL cannot be easily extended to modern wireless systems that use high-order digital modulation schemes such as quadrature amplitude modulation (QAM). This is because the transmitted symbols that are processed by low-order digital modulation (such as the symbols -1 and +1 in BPSK) are linearly superimposed. This linear superimposition does not exist in high-order digital modulation schemes with complex mapping relationships between bits and symbols (such as Gray code).

\subsection{Contributions}

The main contribution of this paper is to develop a novel AirComp FL framework over high-order digital modulation-based wireless systems. %To our best knowledge, \emph{this is the first work that provides a systematic analysis of the integration of digital modulation into an AirComp-based MIMO system for FL performance optimization}. 
Our key contributions include:

\begin{itemize}

    \item We propose a novel AirComp-based MIMO system in which distributed wireless devices utilize high-order modulations to encode their trained local FL parameters into symbols and simultaneously transmit these modulated symbols over unreliable wireless channels to a PS that directly generates the global FL model via its received symbols. However, the introduction of high-order digital modulations leads to the loss of linearity and superpositivity among the symbols transmitted by each device, which in turn affects the convergence of the trained FL model and poses challenges to FL performance. To tackle this issue, the PS and devices must cooperatively adjust the transmit and receive beamforming matrices to capture the nonlinearity relationship among the modulated symbols and improve FL performance. To this end, we formulate this joint transmit and receive beamforming matrix design problem as an optimization problem whose goal is to minimize the FL training loss. 

    \item To solve this problem, we first analytically characterize how the errors introduced by the proposed AirComp system affect FL training loss. Our analysis shows that the introduced errors caused by wireless transmission (i.e., fading and additive white Gaussian noise) and digital post-processing (i.e., digital demodulation) determine the gap between the optimal FL model that the FL targets to converge and the trained FL model. In particular, the errors caused by wireless transmission depend on the channel conditions and the trained FL model parameters. However, the errors caused by digital post-processing depend on the adopted modulation scheme and the number of devices participated in FL training. Hence, to minimize the errors caused by both wireless transmission and digital post-processing, the PS and the devices must dynamically adjust the transmit and receive beamforming matrices based on the adopted modulation scheme, the trained FL model parameters, and channel conditions.
    
    \item To find the optimal transmit and receive beamforming matrices, we first introduce an artificial neural network (ANN)-based algorithm to predict the FL model parameters of all devices since optimizing beamforming matrices requires the information of each trained local model parameter which cannot be obtained by the PS. Then, given the predicted parameters, we derive a closed-form solution of the optimal transmit and receive beamforming matrices based on the adopted modulation scheme and channel conditions that minimize the distance between the received signals of all devices and the predicted parameters in the decision region, which ensures the accuracy for model aggregation and FL performance. We show that the added latency introduced by the ANN inference is marginal compared to the improvement in convergence speed provided by our beamforming optimization.

    \item Numerical evaluation results on real-world machine learning task datasets show that our proposed AirComp-based system can improve the test accuracy by 10\%-30\% compared to the AirComp-based system with analog modulation and BPSK, respectively. 
 
\end{itemize}

The rest of this paper is organized as follows. The system model and problem formulation for the AirComp-based system in FL framework are described in Section II. Section III analyzes the convergence of the designed FL framework and derives a closed-form optimal design of the transmit and receive beamforming matrices based on the analysis. In Section IV, our numerical evaluation is presented and discussed. Finally, conclusions are drawn in Section V.

\section{System Model and Problem Formulation}

\begin{figure*}[t]
\centering
\setlength{\belowcaptionskip}{-0.45cm}
\vspace{-0.1cm}
\includegraphics[width=17cm]{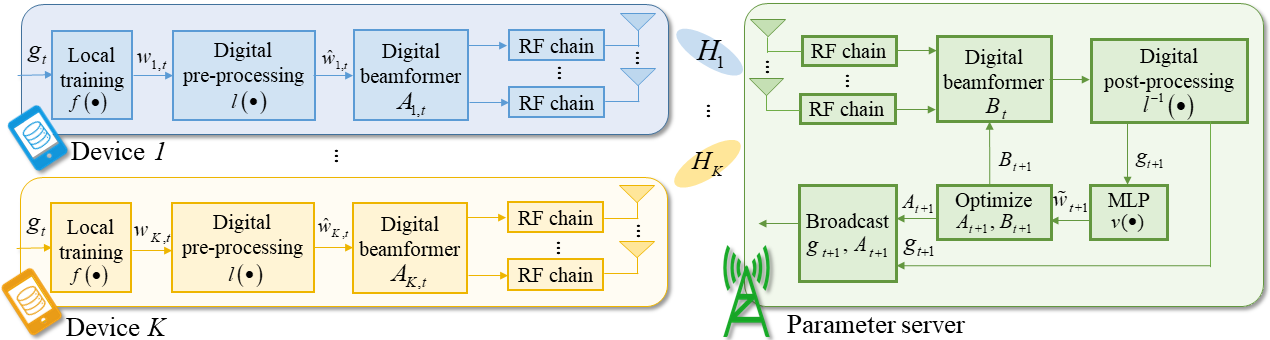}
\centering
\vspace{-0.1cm}
\caption{Illustration of our methodology and system model. An FL algorithm is deployed over multiple devices and one PS in a MIMO communication system. We design the transmit and receive beamforming matrices to optimize the FL training process.}
\vspace{-0.1cm}
\label{system1}
\end{figure*}

We consider an FL system implemented over a cellular network, where $K$ wireless edge devices train their individual machine learning models and send the machine learning parameters to a central PS through a noisy wireless MAC as shown in Fig. \ref{system1}. In the considered model, the PS is equipped with $N_r$ antennas while each device $k$ is equipped with $N_t$ antennas. 

Each device has $N_k$ training data samples and each training data sample $n$ in device $k$ consists of an input feature vector $\bm x_{k,n} \in {\bf R}^{N_{I}\times 1}$ and a corresponding label vector $\bm y_{k,n} \in {\bf R}^{N_{O}\times 1}$ where $N_{I}$ and $N_{O}$ are the dimension of input and output vectors, respectively. Table I provides a summary of the notations used throughout this paper. The objective of the training is to minimize the global loss function over all data samples, which is given by
\begin{equation}\label{eq:FLloss}
\begin{aligned}
F(\bm g)=\mathop {\min }\limits_{\bm g} \frac{1}{N}\sum\limits_{k = 1}^K \sum\limits_{n = 1}^{N_k} f\left(\bm g,\bm x_{k,n},\bm y_{k,n}\right),
\end{aligned}
\end{equation}
where $\bm g \in {\bf R}^{V \times 1}$ is a vector that represents the global FL model of dimension $V$ trained across the devices with $N\!=\!\sum\limits_{k = 1}^{K} \!N_k$ being the total number of training data samples of all devices. $f\left(\bm g,\bm x_{k,n},\bm y_{k,n}\right)$ is the local loss function of each device $k$ with FL model $\bm g$ and data sample $\left(\bm x_{k,n},\bm y_{k,n}\right)$.

\begin{table}[t]
\centering
\renewcommand\arraystretch{1}
\small
%\captionsetup{\font={\normalsize}}
\caption{\normalsize List of Notations}
\vspace{-0.3cm}
\begin{tabular}{ | c || c |}
\hline  Notation & Description \\
\hline  $K$ & Number of devices \\
\hline  $M$ & Adopted modulation order \\
\hline  $N_r$ & Number of antennas on the PS \\
\hline  $N_t$ & Number of antennas on devices \\
\hline  $N_k$ & Number of training data samples on device $k$ \\
\hline  $(\bm x_{k,n},\bm y_{k,n})$ & Training data sample $n$ on device $k$ \\
\hline  $N$ & Number of training data samples of all devices \\
\hline  $\bm g_t$ & Global FL model \\
\hline  $\bm w_{k,t}$ & Local FL model \\
\hline  ${\rm \Delta}{\bm w}_{k,t}$ & Updates of $\bm w_{k,t}$ \\
\hline  $\hat {\bm w}_{k,t}$ & Modulated symbol vector of $\bm w_{k,t}$ \\
\hline  ${\rm \Delta}{\widetilde {\bm w}}_{k,t}$ & Prediction of ${\rm \Delta} {\bm w}_{k,t}$ \\
\hline  ${\rm \Delta}\hat {\bm w}_{k,t}$ & Modulated symbol vector of ${\rm \Delta}{\widetilde {\bm w}}_{k,t}$ \\
\hline  $\bm n_t$ & Additive white Gaussian noise\\
\hline  $\bm A_{k,t}$ & Transmit beamforming matrix \\
\hline  $\bm B_{t}$ & Receive beamforming matrix \\
\hline  $\bm H_{k}$ &  Channel vector between device $k$ and the PS \\
\hline  $P_0$ & Maximal transmit power on device\\
\hline  $\xi$ & Minimum Euclidean distance in decision region \\
\hline  $a_{i}^{\rm I},a_{i}^{\rm Q}$ & Constellation point of symbol $i$  \\
\hline  $\bm a^*$ & Vector of predicted constellation point \\
\hline  $\mathcal{M}$ & Set of all constellation points  \\
\hline
\end{tabular}
\vspace{-0.3cm}
\end{table}

To minimize the global loss function in (\ref{eq:FLloss}) in a distributed manner, each device can update its FL model using its local dataset with a backward propagation (BP) algorithm based on stochastic gradient descent (SGD), which can be expressed as
\begin{equation}\label{eq:update1}
	\begin{aligned}
		\bm w_{k,t}&={\bm g}_{t}-\frac{\lambda}{|\mathcal{N}_{k,t}|} \!\sum\limits_{n\in \mathcal{N}_{k,t}}\!\!\frac{\partial f\left(\bm g,\bm x_{k,n},\bm y_{k,n}\right)}{\bm \partial \bm g},
	\end{aligned}
\end{equation}
where $\lambda$ is the learning rate, $\mathcal{N}_{k,t}$ is the subset of training data samples (i.e., minibatch) selected from device $k$'s training dataset $\mathcal{N}_k$ at iteration $t$ with $|\mathcal{N}_{k,t}|$ being the number of data samples in $\mathcal{N}_{k,t}$, and $\bm w_{k,t}$ is the updated local FL model of device $k$ at iteration $t$. Here, each device must perform $\frac{|\mathcal{N}_{k}|}{|\mathcal{N}_{k,t}|}$ local updates to traverse the entire local training dataset.

Given $\bm w_{k,t}$, distributed devices must simultaneously exchange their model parameters with the PS via bandwidth-limited wireless fading channels for model aggregation. The equation of model aggregation is given by
\begin{equation}\label{eq:localloss}
\begin{aligned}
\bm g_t= \sum\limits_{k = 1}^K \frac{\lvert \mathcal{N}_k \rvert }{N}{\bm w}_{k,t},
\end{aligned}
\end{equation}
where $\lvert \mathcal{N}_k \rvert$ represents the number of data samples in $\mathcal{N}_k$.

To ensure all devices can participate in FL model exchanging via wireless fading channels, each device adopts digital modulation to mitigate wireless fading and the PS adopts beamforming to maximize the number of devices scheduled for FL parameter transmission. Next, we will mathematically introduce the FL training and transmission process integrated with digital modulation in the considered MIMO communication system. In particular, we first introduce our designed digital modulation process that consists of two steps: (i) digital pre-processing at the devices and (ii) digital post-processing at the PS.

%while guarantee the transmission error, 
%(ii) Superposition over the air;

\subsection{Digital Pre-Processing at the Devices}
\label{ssec:preproc}
To transmit ${\bm w}_{k,t}$ over wireless fading channels, each device $k$ leverages digital pre-processing to represent each numerical FL parameter in ${\bm w}_{k,t}$ using a symbol vector, which is
\begin{equation}\label{eq:map}
	\begin{aligned}
		\hat {\bm w}_{k,t}=l\left( {\bm w}_{k,t} \right),
	\end{aligned}
\end{equation}
where $\hat {\bm w}_{k,t} \in {\bf R}^{W}$ is a modulated symbol vector with $W$ being the number of symbols, and $l\left( \cdot \right)$ denotes the digital pre-processing function that combines decimal-to-binary conversion and digital modulation, where the decimal-to-binary conversion is used to represent each numerical FL parameter with a binary coded bit-interleaved vector, and the digital modulation is used to modulate several binary bits as a symbol \cite{code}. For convenience, the modulated signal $\hat {\bm w}_{k,t}$ is normalized (i.e., $|\hat {\bm w}_{k,t}|=1$). We use rectangular $M$-quadrature-amplitude modulation (QAM) for digital modulation and it can be extended to other types of digital modulation schemes. %For the $M$-QAM modulation scheme, every $\sqrt M$ bits in the binary vector are used to represent a symbol in $\hat {\bm w}_{k,t}$ that belongs to a complex constellation $\mathcal{X}$ of size $\sqrt M \times \sqrt M$ (i.e., $\hat {\bm w}_{k,t}=[\hat {w}_{k,t,1}^{\rm I}\hat { w}_{k,t,1}^{\rm Q},\ldots,\hat {w}_{k,t,L}^{\rm I}\hat { w}_{k,t,L}^{\rm Q} ]$ with $\hat {w}_{k,t,L}^{\rm I}$ and $\hat {w}_{k,t,L}^{\rm Q}$ being the in-phase symbol $i$ and quadrature symbol $i$, respectively).
%that is assumed to be normalized by the unit variance (i.e., $\mathbb{E} \left[ \hat {\bm w}_{k,t} \left( \hat {\bm w}_{k,t}\right)^{\rm H} \right]= {\bm I}$)

\iffalse
Motivated by recent standards, rectangular quadrature-amplitude modulation (QAM) is adopted in this work, although extensions to other types of digital modulation are possible. Specifically, for the $M$-QAM modulation, every $\sqrt M$ bits in the binary vector $\widetilde {\bm w}_{k,t}$ are associated via the map function $Q\left(\cdot \right)$ with a symbol in $\hat {\bm w}_{k,t}$ that belongs to a complex constellation $\mathcal{X}$ of size $\sqrt M \times \sqrt M$. The relationship between each symbol in $\widetilde {\bm w}_{k,t}$ and $\hat {\bm w}_{k,t}$ can be expressed by
\begin{equation}\label{eq:QAM}
	\begin{aligned}
		\hat {w}_{k,t}^{(i)}&=Q(\widetilde {\bm w}_{k,t},i)\\
		&=\left\{ \begin{array}{l}
			\sum\limits_{q={\sqrt M}(i-1)+1}^{{\sqrt M}i-\frac{\sqrt M}{2}} 2^{q}{\widetilde w}^{(q)}_{k,t},\qquad {\rm if} \; q \leq \frac{\sqrt M}{2},\\
			\!\!\!\sum\limits_{q={\sqrt M}i-\frac{\sqrt M}{2}+1}^{{\sqrt M}i} \!\!\!j 2^{q-\frac{\sqrt M}{2}}{\widetilde w}^{(q)}_{k,t},\;{\rm otherwise},\\
		\end{array} \right.\
	\end{aligned}
\end{equation}
where $\hat {w}_{k,t}^{(i)}$ is the $i$-th symbol in $\hat {\bm w}_{k,t}=\left[\hat {w}_{k,t}^{(1)},\cdots,\hat {w}_{k,t}^{(L)} \right]$. 
\fi

Given the transmit beamforming matrix $\bm A_{k,t} \in {\bf C}^{N_t \times W}$ and the maximal transmit power $P_0$ at device $k$, the power constraint can be expressed as \cite{XWYB,GK,XGYYL}
\begin{equation}\label{eq:transpower}
	\begin{aligned}
		\mathbb{E}\left( \left| \bm A_{k,t} \hat {\bm w}_{k,t} \right|^2 \right)=\left| \bm A_{k,t} \right|^2\leq P_0.
	\end{aligned}
\end{equation}
where $\mathbb{E}\left( x \right)$ represents the expectation of $x$ and the equality in (\ref{eq:transpower}) is achieved since  the modulated signal is normalized.

\subsection{Post-Processing at the PS}
\label{ssec:postproc}
Considering the multiple access channel property of wireless communication, the received signal at the PS is given by
\begin{equation}\label{eq:wireless1}
	\begin{aligned}
		\bm s_{t}\left(\bm A_t\right)=\sum\limits_{k = 1}^K{\bm H_k} {\bm A_{k,t}} \hat {\bm w}_{k,t}+\bm n_t
	\end{aligned}
\end{equation}
where ${\bm A}_t=[\bm A_{1,t},\cdots,\bm A_{K,t}]$ denotes the transmit beamforming matrices of all devices, ${\bm H_k} \in {\bf C}^{N_r \times N_t}$ denotes the MIMO channel vector for the link from device $k$ to the PS, and $\bm n_t \in {\bf C}^{N_r}$ is the complex additive Gaussian noise with zero mean and identity covariance matrix scaled by the noise power $\sigma^2$, i.e., $\bm n_t \sim \mathcal{C}\mathcal{N}({\bm 0},\sigma^2{\bf I})$.

Since $\bm s_{t}\left(\bm A_t\right)$ is the weighted sum of all users' local FL models, we consider directly generating the global FL model $\bm g_{t+1}$ from $\bm s_{t}(\bm A_{t})$. This is a major difference between the existing works and this work. The digital beamformer output signal can be expressed as
\begin{equation}\label{eq:wireless2}
	\begin{aligned}
		{\bm {\hat s}_{t}}\left(\bm B_{t},\bm A_t\right)={\bm B_{t}}^{\rm H} \bm s_{t}(\bm A_t),
	\end{aligned}
\end{equation}
where $\bm B_t \in {\bf C}^{N_r \times W}$ is the digital receive beamforming matrix.

Given the received symbol vector ${\bm {\hat s}_{t}}\left(\bm B_{t},\bm A_t\right)$, the PS can reconstruct the numerical parameters in global FL model $\bm g_{t+1}$, which can be expressed as
\begin{equation}\label{eq:global}
	\begin{aligned}
		\bm g_{t+1}\left(\bm B_{t},\bm A_t\right)=l^{-1} \left( \bm {\hat s}_{t} \left(\bm B_t,\bm A_t \right) \right),
	\end{aligned}
\end{equation}
where $l^{-1}\left( \cdot \right)$ is the inverse function with respect to $l\left( \cdot \right)$ that combines the binary-to-decimal function and the digital demodulation function. From (\ref{eq:wireless2}) and (\ref{eq:global}), we see that the designed transmit and receive beamforming matrices enable the PS and devices to collaboratively adjust the weights of the transmitted and received signals, thus achieving FL model aggregation.

\subsection{Problem Formulation}

Next, we introduce our optimization problem. Our goal is to minimize the FL training loss by designing the transmit and receive beamforming matrices under the total transmit power constraint of each device, which is formulated as follows:
\begin{equation}\label{eq:max1}
	\begin{split}
		 \mathop {\min}\limits_{\bm B,\bm A} \;\; F\left(\bm g \left(\bm B_T,\bm A_T \right)\right),
	\end{split}
\end{equation}
\vspace{-0.7cm}
\begin{align}\label{c1}
	\setlength{\abovedisplayskip}{-15 pt}
	\setlength{\belowdisplayskip}{-20 pt}
	&\;\;\rm{s.t.}\;\;\scalebox{1}{$ \left| \bm A_{k,t} \right|^2\leq P_0,\forall {k} \in \mathcal{K},\forall {t} \in \mathcal{T}.$}\tag{\theequation a}
	%&~~~~~~~ \left| \left[ \bm U_{t}\right]_{m,n} \right|=1, \forall {m,n}. \tag{\theequation b}
\end{align}
where $\bm A = [\bm A_{1},\ldots,\bm A_{T}]$ and $\bm B = [\bm B_{1},\ldots,\bm B_{T}]$ are the transmit and receive beamforming matrices for all iterations, respectively. $T$ is a constant which is large enough to guarantee the convergence of FL. %In other words, the number of iterations that the FL algorithm requires to converge will not be larger than $T$. (\ref{eq:max1}a) is the constraint of transmit power of each device $k$.

From (\ref{eq:max1}), we can see that the FL training loss $F\left(\bm g \left(\bm B_T,\bm A_T \right)\right)$ depends on the global FL model $g \left(\bm B_T,\bm A_T \right)$ that is trained iteratively. Meanwhile, as shown in (\ref{eq:wireless1}) and (\ref{eq:wireless2}), edge devices and the PS must dynamically adjust $\bm A_{t}$ and $\bm B_{t}$ based on current FL model parameters to minimize the gradient deviation caused by AirComp in the considered MIMO system with digital modulation. However, the PS does not know the gradient vector of each edge device and hence the PS cannot proactively adjust the receive beamforming matrix using traditional optimization algorithms. To tackle this challenge, we propose an ANN-based algorithm that enables the PS to predict the local FL gradient parameters of each device. Based on the predicted local FL model parameters, the PS and edge devices can cooperatively optimize the beamforming matrices to improve the performance of FL. Next, we first mathematically analyze the FL update process in the considered AirComp-based system to capture the relationship between the beamforming matrix design and the FL training loss per iteration. Based on this relationship, we then derive the closed-form solution of optimal $\bm A_t$ and $\bm B_t$ that depends on the predicted FL models achieved by an ANN-based algorithm.

\section{Solution for Problem (\ref{eq:max1})}

%In this section, our goal is to minimize the FL loss function under the transmit power constraint. 
To solve (\ref{eq:max1}), we first analyze the convergence of the considered FL so as to find the relationship between digital beamforming matrices $\bm A_t$, $\bm B_t$, and FL training loss in (\ref{eq:max1}). The analytical result shows that the optimization of beamforming matrices $\bm A_t$ and $\bm B_t$ depends on the FL parameters transmitted by each device. However, the PS does not know these FL parameters since it must determine the beamforming matrices $\bm A_t$ and $\bm B_t$ {\it before} the FL parameter transmission. Therefore, we propose to use neural networks to predict the local FL models of each device and proactively determine the beamforming matrices using these predicted FL parameters.

%To find the tightest bound, we introduced an ANN based algorithm to estimate the local FL models of all devices and then, the optimal solution of beamforming matrices is determined based on the predicted FL model and the derived expected improvement of FL training loss.

\subsection{Convergence Analysis of Designed FL}

We first analyze the convergence of the considered FL system. Since the update of the global FL model depends on the instantaneous signal-to-interference-plus-noise ratio (SINR) affected by the digital beamforming matrices $\bm A_t$ and $\bm B_t$, we can only analyze the expected convergence rate of FL. To analyze the expected convergence rate of FL, we first assume that a) the loss function $F(\bm g)$ is $L-$smooth with the Lipschitz constant $L>0$, b) $F(\bm g)$ is strongly convex with positive parameter $\mu$, c) $F(\bm g)$ is twice-continuously differentiable, and d) $\left\| \nabla f( {\bm g}_{t},\bm x_{kn},\bm y_{kn})\right\|^2 \!\leq\! \zeta_1 \!+\! \zeta_2\left\| \nabla F({\bm g}_{t})\right\|^2,$  as done in \cite{APP}.
% \begin{itemize}
% 	\item \textbf{Assumption 1}: The loss function $F(x)$ is $L-$smooth with the Lipschitz constant $L>0$, such that
% 	\begin{equation}\label{eq:Assumption1}
% 		\begin{aligned}
% 			||\nabla F(x)\!-\!\nabla F(y)||\!\leq\! L||x-y||.
% 		\end{aligned}
% 	\end{equation}
% 	\item \textbf{Assumption 2}: The loss function $F(x)$ is strongly convex with positive parameter $\mu$, such that
% 	\begin{equation}\label{eq:Assumption2}
% 		\begin{aligned}
% 			F(\bm g_{t+1})\geq F(\bm g_{t})+(\bm g_{t+1}\!-\!\bm g_{t})^T \nabla  F(\bm g_{t})\!+\!\frac{\mu}{2}||\bm g_{t+1}\!-\!\bm g_{t} ||.
% 		\end{aligned}
% 	\end{equation}
% 	\item \textbf{Assumption 3}: The loss function $F(x)$ is twice-continuously differentiable. Based on (\ref{eq:Assumption1}) and (\ref{eq:Assumption2}), we have
% 	\begin{equation}\label{eq:Assumption3}
% 		\begin{aligned}
% 			\mu {\bm I} \! \preceq\! \nabla^2 \!F({\bm g}_{t},\!\bm x_{kn},\bm y_{kn})\! \preceq\! L {\bm I}.
% 		\end{aligned}
% 	\end{equation}
% 	\item We also assume that 
% 	\begin{equation}\label{eq:Assumption4}
% 		\begin{aligned}
% 			\left\| \nabla f( {\bm g}_{t},\bm x_{kn},\bm y_{kn})\right\|^2 \!\leq\! \zeta_1 \!+\! \zeta_2\left\| \nabla F({\bm g}_{t})\right\|^2,
% 		\end{aligned}
% 	\end{equation}
% 	where $F({\bm g}_{t})\!=\!\frac{1}{N} \! \sum\limits_{k = 1}^K \!\sum\limits_{n = 1}^{N_k} \!f\!\left(\bm g_t,\bm x_{k,n},\bm y_{k,n}\right)$.
% \end{itemize}
These assumptions can be satisfied by several widely used loss functions such as mean squared error, logistic regression, and cross entropy. Based on these assumptions, next, we first derive the upper bound of the FL training loss at one FL training step. The expected convergence rate of the designed FL algorithm can now be obtained by the following theorem.
\begin{theorem}
{\rm Given the optimal global FL model $\bm g^*$, the current global FL model $\bm g_t$, the transmit beamforming matrix $\bm A_{t}$, and the receive beamforming matrix $\bm B_{t}$, $\mathbb{E}\left(F\left(\bm g_{t+1}(\bm A_{t},\bm B_{t})\right)-F\left(\bm g^*\right)\right)$ can be upper bounded as
\begin{equation}\label{eq:A0}
\begin{split}
&\mathbb{E}\left(F\left({\bm g}_{t+1}(\bm A_{t},\bm B_{t})\right)-F\left({\bm g}^*\right)\right)\\
\leqslant&  \mathbb{E}\left(F\left({\bm g}_{t}\right)-F\left({\bm g}^*\right)\right)
-\frac{1}{2L}\left\|\nabla F\left({\bm g}_{t}\right)\right\|^2\\
&+\frac{1}{2L}\mathbb{E}\left(\left\| \bm e_t\right\|+\left\|\hat{\bm e_t}(\bm A_{t},\bm B_{t})\right\|\right)^2,
\end{split}
\end{equation}
where 
\begin{equation}\label{eq:A00}
\begin{split}
\bm e_t=&\left\| \frac{\sum\limits_{k = 1}^{K}  \sum\limits_{n\in \mathcal{N}_{k,t}} \nabla f\left(\bm g_t,\bm x_{n,k}, \bm y_{n,k}\right) }{\sum\limits_{k = 1}^{K} |\mathcal{N}_{k,t}|} \right. \\
&\left. -l^{-1}\left(\frac{\sum\limits_{k = 1}^{K} l \left( \sum\limits_{n\in \mathcal{N}_{k,t}}  \nabla f\left(\bm g_t,\bm x_{n,k}, \bm y_{n,k}\right) \right)}{\sum\limits_{k = 1}^{K} |\mathcal{N}_{k,t}|}\right)\right\| + \Upsilon
\end{split}
\end{equation}
with the first term being the gradient trained by SGD , the second term being the gradient demodulated from a sum of all selected devices' symbols, and $\Upsilon$ being the variance bound introduced by random sampling of mini-batches of all devices. In particular, $\Upsilon$ depends on the variance of the local updated gradient in minibatch SGD, which is \cite{NMJ,ZAA,LJD}
\begin{equation}\label{eq:update2}
\begin{aligned}
\mathbb{E}|| \widehat{\bm w}_{k,t}- \bm w_{k,t} ||=\frac{\sigma^2_k}{|\mathcal{N}_{k}|},
\end{aligned}
\end{equation}
where $\widehat{\bm w}_{k,t}$ is the optimal updated local model without variance and $\sigma^2_k$ is the variance of the local gradient. Given $\frac{\sigma^2_k}{|\mathcal{N}_{k}|}$ on each device $k$, the relationship between $\Upsilon$ and $\frac{\sigma^2_k}{|\mathcal{N}_{k}|}$ is
\begin{equation}\label{eq:update3}
\begin{aligned}
\Upsilon &= \sum\limits_{k = 1}^{K} \frac{\sigma^2_k}{|\mathcal{N}_{k}|},
\end{aligned}
\end{equation}}
and
\begin{equation}\label{eq:A02}
\begin{split}
&\hat {\bm e_t}(\bm A_{t},\bm B_{t})= l^{-1}\left(\frac{\sum\limits_{k = 1}^{K} l \left( \sum\limits_{n\in \mathcal{N}_{k,t}}  \nabla f\left(\bm g_t,\bm x_{n,k}, \bm y_{n,k}\right) \right)}{\sum\limits_{k = 1}^{K} |\mathcal{N}_{k,t}|}\right)\\
&-l^{-1} \left( \!\frac{\bm B_t \left(\sum\limits_{k = 1}^{K} \bm H_k \bm A_{k,t} l \left( \sum\limits_{n\in \mathcal{N}_{k,t}} \! \!\!\! \nabla f\left(\bm g_t, \bm x_{n,k}, \bm y_{n,k}\right)\!\right) \!+\!\bm n_t\right)}{\sum\limits_{k = 1}^{K} |\mathcal{N}_{k,t}|} \!\right). 
\end{split}
\end{equation}
%is a gap between the gradient demodulated from an averaged superpositioned signal withot channel fading or noise and the gradient demodulated from the received signal (i.e., a gradient gap caused by the errors of digital demodulation).}
\end{theorem}
\begin{IEEEproof}See Appendix~\ref{app:proof}.
\end{IEEEproof}

From Theorem 1, we see that, since $\bm e_t$ does not depend on $\bm A_t$ or $\bm B_t$, the optimization of the digital beamforming matrices cannot minimize $\bm e_t$. In consequence, we can only minimize $\left\| \bm {\hat e}_t \right\|$ to decrease the gap between the FL training loss at iteration $t+1$ and the optimal FL training loss (i.e., $\mathbb{E}\left(F\left({\bm g}_{t+1}\right)-F\left({\bm g}^*\right)\right)$). Thus, problem (\ref{eq:max1}) can be rewritten as
%In addition, Theorem 1 also shows that the performance of the global FL model $F\left(\bm g_{t+1}\right)$ 
%{\color{blue}depends on only the data distribution, the modulation order $M$, and the number of devices $K$}
\begin{equation}\label{eq:max2}
	\begin{split}
		 \mathop {\min}\limits_{\bm B_t,\bm A_t} \;\; 
		 \left\|\hat{\bm e_t}\right\|^2
	\end{split}
\end{equation}
\vspace{-0.4cm}
\begin{align}\label{c1}
	\setlength{\abovedisplayskip}{-15 pt}
	\setlength{\belowdisplayskip}{-20 pt}
	&\;\;\rm{s.t.}\;\;\scalebox{1}{$ \left| \bm A_{k,t} \right|^2\leq P_0,\forall {k} \in \mathcal{K},\forall {t} \in \mathcal{T}.$}\tag{\theequation a}
\end{align}

To minimize $\left\| \bm {\hat e}_t \right\|$ in (\ref{eq:max2}), the PS and edge devices must obtain the information of MIMO channel vector $\bm H_k$ as well as the trained gradients $l \left( \sum\limits_{n\in \mathcal{N}_{k,t}}  \nabla f\left(\bm g_t,\bm x_{n,k}, \bm y_{n,k}\right) \right)$ so as to adjust $\bm A_t$ and $\bm B_t$. However, the trained FL gradients ${\rm \Delta}{\bm w}_{k,t}=\sum\limits_{n\in \mathcal{N}_{k,t}}  \nabla f\left(\bm g_t,\bm x_{n,k}, \bm y_{n,k}\right)$ cannot be obtained by the PS before edge devices sending FL model parameters. Hence, the PS must predict ${\rm \Delta}{\bm w}_{k,t}$ for optimizing $\bm A_t$ and $\bm B_t$ and minimizing $\left\| \bm {\hat e}_t \right\|$.

\subsection{Prediction of Local FL Models}
\label{ssec:mlp}
Next, we explain the use of neural networks to predict the local FL model updates of all devices at the PS. Formally, the task is to predict parameters of device $k$'s FL model update at time $t$, i.e., ${\rm \Delta}{\bm w}_{k,t} \in {\bf R}^{V \times 1}$, from the global aggregation available at the PS at time $t-1$, i.e., ${\bm g}_{t-1} \in {\bf R}^{V \times 1}$. This is considered a regression task since the values of the FL model are continuous. Therefore, we propose to use multilayer perceptrons (MLPs), a standard type of artificial neural network (ANN) for handling regression tasks \cite{book}. 

Our proposed MLP-based prediction algorithm consists of three layers: (a) input layer, (b) a single hidden layer, and (c) output layer. These components are defined as follows:
\begin{itemize}
\item {\emph {Input layer:}} The input to the MLP is a vector ${\bm g}'_{t-1}\in {\bf R}^{V' \times 1}$ that represents the previous aggregated results of the FL parameters being predicted. This is a subset of ${\bm g}_{t-1}\in {\bf R}^{V \times 1}$, where the number of parameters $V' \leq V$ is adjustable. Since the total parameter dimension $V$ may be large (e.g., for the CIFAR-10 experiments in Sec.~\ref{sec:exp}, $V = 1.17 \times 10^7$), selecting a smaller number $V'$ to predict can lead to complexity reduction advantages. As we mentioned in (\ref{eq:wireless1}), all devices are able to connect with the PS so as to provide the input information for the MLP to predict the local FL models for next iteration.

%This MLP output is combined with other $V - V'$ parameters in ${\rm \Delta}{\bm w}_{k,t}$ (i.e., those that are not part of the MLP) to form ${\rm \Delta}{\widetilde {\bm w}}_{k,t} \in {\bf R}^{V \times 1}$. The PS will adjust the transmit and receive beamforming matrices to minimize (\ref{eq:max2}) based on ${\rm \Delta}{\widetilde {\bm w}}_{k,t}$. 
\item {\emph {Output layer:}} The output of the MLP is a vector ${\rm \Delta}{\widetilde {\bm w}'}_{k,t} \in {\bf R}^{V' \times 1}$ that represents prediction of the $V'$ parameters in device $k$'s local FL model update for the current iteration $t$. The resulting estimate ${\rm \Delta}{\widetilde {\bm w}}_{k,t} \in {\bf R}^{V \times 1}$ of ${\rm \Delta}{\bm w}_{k,t}$ used in the beamforming design in Section III-C is then a combination of (a) the $V'$ parameter predictions ${\rm \Delta}{\widetilde {\bm w}'}_{k,t}$ that the ANN has made, and (b) the other $V - V'$ parameters in ${\bm g}_{t-1}$ that are not part of the ANN.

%ensure the minimization of gradient deviation during AirComp transmission in the considered MIMO system with digital modulation.
	
\item {\emph {Single hidden layer:}} We employ a single hidden layer of dimension $D$ to learn the nonlinear relationships between the input ${\bm g}'_{t-1}$ and the output ${\rm \Delta}{\widetilde {\bm w}}'_{k,t}$. The weight matrix between the input vector and the neurons in the hidden layer for device $k$ is denoted ${\bm v}^{\rm in}_k \in {\bf R}^{V' \times D}$. Meanwhile, the weight matrix between the neurons in the hidden layer and the output vector is denoted ${\bm v}^{\rm out}_k \in {\bf R}^{D \times V'}$.
\end{itemize}
Based on this, the states of the neurons in the hidden layer are given by
\begin{equation}\label{eq:hidden}
	\begin{aligned}
		{\bm v}_{k,t} = \sigma \left({\bm v}_k^{\rm in} {\bm g}'_{t-1} + {\bm b_k^v}\right),
	\end{aligned}
\end{equation}
where $\sigma \left(x \right)=\frac{2}{1+{\rm exp}\left(-2x\right)}-1$ is a sigmoidal activation and ${\bm b_k^v} \in {\bf R}^{D' \times 1}$ is the bias vector. Then, the output of the MLP is given by
\begin{equation}\label{eq:output}
	\begin{aligned}
		{\rm \Delta}{\widetilde {\bm w}'}_{k,t}= {\bm v}_k^{\rm out} \bm v_{k,t} +{\bm b_k^o},
	\end{aligned}
\end{equation}
where ${\bm b_k^o} \in {\bf R}^{V' \times 1} $ is another bias vector.

The MLP is trained through an online gradient descent method in parallel with the FL model updating. However, in the considered model, the PS only has access to the value of $\bm g_t$ that is directly demodulated from the received signal from all devices. Hence, the PS and the devices must exchange information to train the MLP for each device cooperatively. In particular, at each iteration, device $k$ first calculates its local update ${\rm \Delta}\bm w_{k,t}$ and $\bm g_{t-1}$ received from the PS, and extracts ${\rm \Delta}\bm w_{k,t}' \in {\bf R}^{V' \times 1}$ from ${\rm \Delta}\bm w_{k,t}$. Then, it calculates the gradient of the MLP parameters with respect to the MLP's prediction of ${\rm \Delta}\bm w_{k,t}'$ from $\bm g'_{t-1}$. Finally, device $k$ transmits these gradients to the PS, which updates the MLP parameters accordingly. We will see in Sec.~\ref{sec:exp} that the per-iteration latency introduced by this MLP procedure is overcome by the reduction in the number of training rounds needed for convergence, so long as the size of the MLP is limited.
%Since each device only needs to transmit its training loss, the cost for information exchange can be ignored compared with the transmission of DNN model parameters in FL training.

\subsection{Optimization of the Beamforming Matrices}

\begin{figure}[t]
	\centering
	\setlength{\belowcaptionskip}{-0.45cm}
	\vspace{-0.1cm}
	\includegraphics[width=8cm]{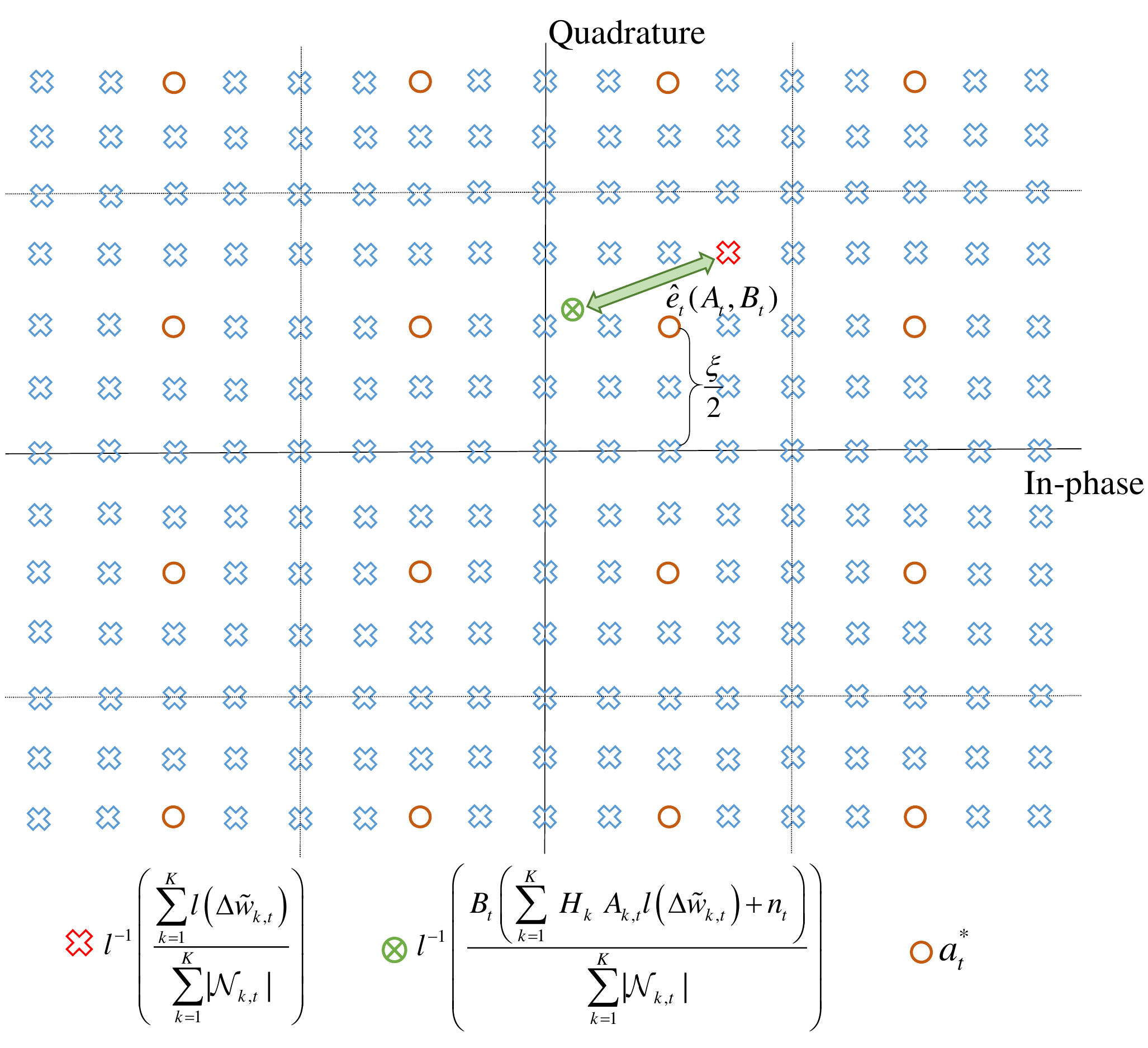}
	\centering
	\vspace{-0.2cm}
	\caption{An example of 16-QAM constellation at the PS with 4 devices.}
	\vspace{-0.2cm}
	\label{system2}
 \vspace{-0.2cm}
\end{figure}

Having the predicted local FL model updates ${\rm \Delta}{\widetilde {\bm w}}_{k,t}$, the PS can optimize the beamforming matrices $\bm A_t$ and $\bm B_{t}$ to solve Problem (\ref{eq:max2}). Substituting ${\rm \Delta}{\widetilde {\bm w}}_{k,t}$, (\ref{eq:wireless1}), and (\ref{eq:wireless2}) into (\ref{eq:max2}), we have
\begin{equation}\label{eq:max3}
\begin{split}
\mathop {\min}\limits_{\bm B_t,\bm A_t} \;\; 
&\left\|l^{-1}\left(\frac{\sum\limits_{k = 1}^{K} l \left( {\rm \Delta}{\widetilde {\bm w}}_{k,t} \right)}{\sum\limits_{k = 1}^{K} |\mathcal{N}_{k,t}|}\right) \right.\\
&\left. -l^{-1} \left( \frac{\bm B_t \left(\sum\limits_{k = 1}^{K} \bm H_k \bm A_{k,t} l \left( {\rm \Delta}{\widetilde {\bm w}}_{k,t}\right) +\bm n_t\right)}{\sum\limits_{k = 1}^{K} |\mathcal{N}_{k,t}|} \right)\right\|^2
\end{split}
\end{equation}
\vspace{-0.2cm}
\begin{align}\label{c1}
	\setlength{\abovedisplayskip}{-15 pt}
	\setlength{\belowdisplayskip}{-20 pt}
	&\;\;\rm{s.t.}\;\;\scalebox{1}{$ \left| \bm A_{k,t} \right|^2\leq P_0,\forall {k} \in \mathcal{K},\forall {t} \in \mathcal{T}.$}\tag{\theequation a}
\end{align}
In (\ref{eq:max3}), $l^{-1}\left(\frac{\sum\limits_{k = 1}^{K} l \left( {\rm \Delta}{\widetilde {\bm w}}_{k,t} \right)}{\sum\limits_{k = 1}^{K} |\mathcal{N}_{k,t}|}\right)$ is independent of $\bm A_t$ and $\bm B_t$ and can be regarded as a constant. However, the existence of the inverse function $l^{-1}(\cdot)$ defined in (\ref{eq:global}) significantly increases the complexity for solving (\ref{eq:max3}). Considering $l^{-1}(\cdot)$ that is used to demodulate the symbols into numerical FL parameters, the minimization of the gap between $l^{-1}\left(\frac{\sum\limits_{k = 1}^{K} l \left( {\rm \Delta}{\widetilde {\bm w}}_{k,t} \right)}{\sum\limits_{k = 1}^{K} |\mathcal{N}_{k,t}|}\right)$ and $l^{-1} \left( \frac{\bm B_t \left(\sum\limits_{k = 1}^{K} \bm H_k \bm A_{k,t} l \left( {\rm \Delta}{\widetilde {\bm w}}_{k,t}\right) +\bm n_t\right)}{\sum\limits_{k = 1}^{K} |\mathcal{N}_{k,t}|} \right)$ is equivalent to minimize the distance between $\frac{\sum\limits_{k = 1}^{K} l \left( {\rm \Delta}{\widetilde {\bm w}}_{k,t} \right)}{\sum\limits_{k = 1}^{K} |\mathcal{N}_{k,t}|}$ and $ \frac{\bm B_t \left(\sum\limits_{k = 1}^{K} \bm H_k \bm A_{k,t} l \left( {\rm \Delta}{\widetilde {\bm w}}_{k,t}\right) +\bm n_t\right)}{\sum\limits_{k = 1}^{K} |\mathcal{N}_{k,t}|}$ in the decision region of digital demodulation, as shown in Fig. \ref{system2}. To this end, in this section, we first derive the position of $\frac{\sum\limits_{k = 1}^{K} l \left( {\rm \Delta}{\widetilde {\bm w}}_{k,t} \right)}{\sum\limits_{k = 1}^{K} |\mathcal{N}_{k,t}|}$ in the decision region and remove $l^{-1}(\cdot)$ from (\ref{eq:max3}) for simplification. Then, we present a closed-form optimal design of the transmit and receive beamforming matrices.

%However, the existence of the inverse function $l^{-1}(\cdot)$ that is used to demodulate the symbols into numerical FL parameters significantly increases the complexity for solving (\ref{eq:max3}). Hence, in this section, we first analyze the mathematical relationship between $l^{-1}(x)$ and $x$ to remove $l^{-1}(\cdot)$ in (\ref{eq:max3}) for simplification. 

Given ${\rm \Delta}{\widetilde {\bm w}}_{k,t}$ and the digital pre-processing function $l(\cdot)$ defined in (\ref{eq:map}), the modulated symbol vector ${\rm \Delta} \hat {\bm w}_{k,t}=l \left( {\rm \Delta}{\widetilde {\bm w}}_{k,t} \right)=[{\rm \Delta} \hat {w}_{k,t,1}^{\rm I} {\rm \Delta}\hat {w}_{k,t,1}^{\rm Q},\ldots,{\rm \Delta} \hat {w}_{k,t,L}^{\rm I} {\rm \Delta}\hat {w}_{k,t,L}^{\rm Q}]$ can be obtained where ${\rm \Delta} \hat {w}_{k,t,i}^{\rm I}$ and ${\rm \Delta} \hat {w}_{k,t,i}^{\rm Q}$ are the $i$-th in-phase and quadrature symbols modulated by ${\rm \Delta}{\widetilde {\bm w}}_{k,t}$, respectively. Since in-phase and quadrature-phase symbols that have vertical and horizontal decision regions are mutually independent, the value of $l^{-1}\left(\frac{\sum\limits_{k = 1}^{K} {\rm \Delta} \hat {\bm w}_{k,t}}{\sum\limits_{k = 1}^{K} |\mathcal{N}_{k,t}|}\right)$ can be obtained via individually analyzing the decision region of each in-phase and quadrature-phase symbols which are
\begin{equation}\label{eq:Euclid1}
\begin{split}
\left| \frac{1}{\sum\limits_{k=1}^K |\mathcal{N}_{k,t}|}{\sum\limits_{k=1}^K {\rm \Delta} \hat {w}_{k,t,i}^{\rm I}} -a_{i}^{\rm I}\right|\leqslant \frac{\xi }{2},\\
 \left| \frac{1}{\sum\limits_{k=1}^K |\mathcal{N}_{k,t}|}{\sum\limits_{k=1}^K {\rm \Delta} \hat {w}_{k,t,i}^{\rm Q}} -a_{i}^{\rm Q}\right|\leqslant \frac{\xi }{2},
\end{split} 
\end{equation}
where $a_{i}^{\rm I},a_{i}^{\rm Q}  \in \mathcal{M} =\left\{ \frac{1-\sqrt{M}}{2}\xi,\frac{3-\sqrt{M}}{2}\xi,\ldots,\frac{\sqrt{M}-1}{2}\xi \right\}$ are the constellation points in the decision region with $\mathcal{M}$ being the set of all constellation points. $\xi=\sqrt{\frac{4P_0}{\left(\sqrt{M}-1\right)^2}}$ is the minimum Euclidean distance between two constellation points. Using (\ref{eq:Euclid1}), $a_{i}^{\rm I}$ and $a_{i}^{\rm Q}$ are given by
\begin{equation}\label{eq:Euclid2}
\begin{split}
&a_{t,i}^{\rm I}=\left\{ x \Big| -\frac{\xi}{2}+ \frac{\sum\limits_{k=1}^K {\rm \Delta} \hat {w}_{k,t,i}^{\rm I} }{\sum\limits_{k=1}^K |\mathcal{N}_{k,t}|} \leqslant x \leqslant \frac{\xi}{2}+ \frac{\sum\limits_{k=1}^K {\rm \Delta} \hat {w}_{k,t,i}^{\rm I} }{\sum\limits_{k=1}^K |\mathcal{N}_{k,t}|} \cap \mathcal{M}  \right\}
\end{split}
\end{equation}
and
\begin{equation}\label{eq:Euclid3}
\begin{split}
&a_{t,i}^{\rm Q}=\left\{ x \Big| -\frac{\xi}{2}+ \frac{\sum\limits_{k=1}^K {\rm \Delta} \hat {w}_{k,t,i}^{\rm Q} }{\sum\limits_{k=1}^K |\mathcal{N}_{k,t}|} \leqslant x \leqslant \frac{\xi}{2}+ \frac{\sum\limits_{k=1}^K {\rm \Delta} \hat {w}_{k,t,i}^{\rm Q} }{\sum\limits_{k=1}^K |\mathcal{N}_{k,t}|} \cap \mathcal{M} \right\}.
	\end{split}
\end{equation}

Given $\bm a_t^{*}=[a_{t,1}^{\rm I}a_{t,1}^{\rm Q},\ldots,a_{t,W}^{\rm I}a_{t,W}^{\rm Q}]$, problem (\ref{eq:max3}) can be rewritten as
\vspace{0.1cm}
\begin{equation}\label{eq:max4}
	\begin{split}
		 \mathop {\min}\limits_{\bm B_t,\bm A_t} \;\; 
		 \left\|\bm a_t^{*}-{\bm B_{t}}^{\rm H}\sum\limits_{k = 1}^K{\bm H_k} {\bm A_{k,t}}  {\rm \Delta}\hat {\bm w}_{k,t}-{\bm B_{t}}^{\rm H}\bm n_t\right\|^2
	\end{split}
\end{equation}
\vspace{-0.3cm}
\begin{align}\label{c1}
	\setlength{\abovedisplayskip}{-15 pt}
	\setlength{\belowdisplayskip}{-20 pt}
	&\;\;\rm{s.t.}\;\;\scalebox{1}{$ \left\| \bm A_{k,t} \right\|^2\leq P_0,\forall {k} \in \mathcal{K},\forall {t} \in \mathcal{T}.$}\tag{\theequation a}
\end{align}
where ${\rm \Delta}\hat {\bm w}_{k,t}=l\left( {\rm \Delta}{\widetilde {\bm w}}_{k,t} \right)$ is a modulated symbol vector of ${\rm \Delta}{\widetilde {\bm w}}_{k,t}$. Problem (\ref{eq:max4}) can be solved by an iterative optimization algorithm. In particular, to solve problem (\ref{eq:max4}), we first fix $\bm B_t$, then the objective functions and constraints with respect to $\bm A_t$ are convex and can be optimally solved by using a dual method \cite{book1}. Similarly, given $\bm A_t$, problem (\ref{eq:max4}) is minimized as $\bm B^*_t=\left(\frac{\bm a_t^{*}}{\sum\limits_{k = 1}^K{\bm H_k} {\bm A^*_{k,t}}  {\rm \Delta}\hat {\bm w}_{k,t}}\right)^{\rm H}$.
%that first optimizes transmit beamforming matrix $\bm A_t$ with fixed receive beamforming matrix $\bm B_t$, and then finds the optimal receive beamforming matrix $\bm B_t$ with optimized transmit beamforming matrix $\bm A_t$.

\subsection{Implementation and Complexity}
Next, we discuss the implementation and complexity of the designed FL algorithm. With regards to the implementation of the proposed algorithm, the PS must a) use MLPs to predict the devices' local FL model update, and b) design the transmit and receive beamforming matrices based on the predicted updates. % and {\color{blue}c) satisfy the synchronization requirement in the proposed AirComp framework.} 
To train the MLPs that are used for the predictions of devices' local FL model update, the PS will use the global FL model $\bm g_{t-1}$ that is directly reconstructed from the received symbol vector ${\bm {\hat s}_{t-1}}$ at iteration $t-1$. Additionally, the PS needs the gradients of the MLP parameters ${\bm v}_k^{\rm in}, {\bm v}_k^{\rm out}$ calculated on the device. To design the optimal transmit and receive beamforming matrices, the PS requires the maximal transmit power $P_0$ and the MIMO channel vector $\bm H_k$ of each device $k$. We ignore the overhead of each device transmitting $P_0$ to the PS since it is a scalar. With regards to $\bm H_k$, the PS can use channel estimation methods to learn $\bm H_k$ over each uplink channel so as to design optimal transmit and receive beamforming matrices. 

To satisfy the synchronization requirement in the proposed AirComp framework, it is necessary to incorporate clock synchronization for precise control of information transmission timing and channel state information (CSI) feedback, for accurate estimation of transmission delay. For clock synchronization, the PS can broadcast a shared block to all devices to achieve time calibration, as in previous studies \cite{OHDM,ZGYL,FCF}. For accurate estimation of transmission delay, the PS can obtain the CSI directly from the uplink pilot signals transmitted from devices, as done in previous studies \cite{AFV,MZ,XPXD}.

%Under the conditions of stringent clock synchronization and accurate estimation of transmission delay, each device can choose the transmission timing to ensure that the signals arrive at the PS synchronously. In addition, synchronization is not only required for AirComp, but also for other communication technologies, such as orthogonal frequency division multiplexing. Therefore, synchronization does not impose additional communication overhead on the proposed AirComp-based systems.

We identify two components of our algorithm that could potentially impact complexity and latency: (1) the MLP and (2) optimizing $\bm A_t$ and $\bm B_t$. The training complexity of the MLP can be made small compared to the complexity of FL training. Specifically, the computational complexity of the MLP lies in the size of input $\bm g'_{t-1}$ and output ${\rm \Delta}{\widetilde {\bm w}}'_{k,t}$, as well as the number of the neurons in the hidden layer. As discussed in Sec.~\ref{ssec:mlp}, the sizes of $\bm g'_{t-1}$ and ${\rm \Delta}{\widetilde {\bm w}}'_{k,t}$ are both $V'$, while the number of neurons in the hidden layer is $D$. Hence, the computational complexity is ${\mathcal O}(2 V' D)$, which implies that the training complexity of the MLP can be controlled directly based on limiting the size of $V'$ and $D$. In Sec.~\ref{sec:exp}, for the ML task, we will set MLP to predict parameters for the last layer of the neural network, so that $V' \ll V$. We can further select $D$ such that $2 V' D \ll V$, i.e., so the MLP is substantially smaller than the FL model.
%The relatively low-training complexity of the MLP also implies that the additional computational latency introduced by MLP inference is relatively minor. Communication of the model parameters between the clients and the PS is much more time consuming than forward passing the signal through an ANN at the PS.

Regarding the optimization of $\bm A_t$ and $\bm B_t$, the complexity scales in the number of iterations required for the solver to converge. For finding optimal $\bm A_t$ and $\bm B_t$, problem (\ref{eq:max4}) can be solved by a traditional augmented Lagrangian method that approaches the optimal solution via alternating updating $\bm A_t$, $\bm B_t$, and the Lagrangian multiplier vector. The introduced Lagrangian multiplier vector consists of $K$ constraints, where $K$ is the number of devices in the considered FL framework. Hence, the PS is required to sequentially update $K$ Lagrangian multipliers, $\bm A_t=[\bm A_{1,t},\cdots,\bm A_{K,t}]$, and $\bm B_t$ at each iteration. Letting $L_O$ be the number of iterations until the augmented Lagrangian method converges, the complexity is $\mathcal{O}(L_O K^2)$.

%This is because the complexity of training an MLP depends on the training data samples and the number of devices. 

\begin{algorithm}[t]
%\footnotesize
\caption{Proposed FL Over AirComp-based System}
\begin{algorithmic}[1]
\STATE \textbf{Init:} Global FL model $\bm g_0$, beamforming metrics $\bm A_{0}$ and $\bm B_{0}$, MIMO channel matrix $\bm H$.
\FOR {iterations $t=0,1,\cdots,T$}
\FOR {$k \in \left\{1, 2, \cdots, K \right\}$ in parallel over $K$ devices}
\STATE Each device calculates and returns $\bm{w}_{k,t}$ based on local dataset and $\bm g_t$ in (\ref{eq:update1}).
\STATE Each device leverages digital pre-processing to modulate each model parameter into a symbol.
\STATE Each device sends the symbol vector $\bm{\hat w}_{k,k}$ to the PS using the optimized transmit beamforming matrix $\bm A_{k,t}$.
\ENDFOR
\STATE The PS directly demodulates the global FL model $\bm g_{t+1}$ from the received superpositioned signal using (\ref{eq:global}).
\STATE The PS predicts the local FL model $\bm{\hat w}_{k,t+1}$ of each device based on demodulated $\bm g_{t+1}$ using trained ANNs.
\STATE The PS proactively adjusts the transmit and receive beamforming matrices using the augmented Lagrangian method and broadcast the transmit beamforming matrix $\bm A_{k,t+1}$ to each device $k$.
\ENDFOR
\end{algorithmic}
\label{algorithm_2}
\end{algorithm}

\section{Experimental Results and Discussion}
\label{sec:exp}
\subsection{Simulation Setup}
We consider a circular network area with a radius $r = 1500$ m with one PS at its center serving $K = 20$ uniformly distributed devices. In particular, the PS allocates 64 subcarriers to all devices and the bandwidth of each subcarrier is 15 kHz. By default, the channels between the PS and devices are modeled as independent and identically distributed with Rayleigh fading and a pathloss exponent of $\beta = 2$. The other default settings of parameters used in simulations are listed in Table II.

For comparison purposes, we consider several baseline AirComp FL frameworks: analog modulation, BPSK modulation, 64 QAM without the assistance of MLP, 64 QAM replacing the MLP with a tabular approach, and an ideal upper bound of 64 QAM over noiseless channels. These considered baselines are detailed as follows:
\begin{itemize}

    \item The analog AirComp FL framework, from \cite{SJY5}, enables each device to use digital beamforming and analog modulation for transmitting FL parameters over wireless fading channels (labeled ``Analog FL'' in plots).

    \item The BPSK AirComp FL framework, from \cite{XLR8}, enables each device to quantize its trained FL model parameters into one bit and use digital beamforming and BPSK for transmitting quantized FL parameters over wireless fading channels (labeled ``BPSK FL'' in plots).

    \item An AirComp FL framework without the assistance of MLP enables the devices to quantize their trained FL model parameters into 6 bits and directly use digital beamforming and 64 QAM for transmitting quantized FL parameters over wireless fading channels (labeled ``Proposed method without MLP'' in plots).

    \item An AirComp FL framework considering noiseless channels enables the devices to quantize their trained FL model parameters into 6 bits and use digital beamforming and 64 QAM for quantized FL parameter transmission over noiseless channels. Then, the PS uses the receive beamforming matrix optimized by MLP to adjust the decision region of the received superimposed signals to minimize the transmission error (labeled ``Proposed FL over noiseless channels'' in plots).

    \item An AirComp FL framework where, as in our method, the devices quantize their trained FL model parameters into 6 bits and directly use digital beamforming and 64 QAM for transmitting quantized FL parameters. However, ChannelComp from \cite{SJC} is employed to optimize the transmission error instead of the ANN (labeled ``Proposed method with ChannelComp'' in plots).
\end{itemize}

% MNIST dataset \cite{MNIST}, 
%MNIST, we adopt a fully-connected neural network (FNN) that consists of two full-connection layers with 7840 (=28$\times$28$\times$10) model parameters. For 
The Fashion-MNIST dataset \cite{Fashion-MNIST} and CIFAR-10 dataset \cite{CIA10} are used as ML tasks in our performance evaluation. For Fashion-MNIST, each local FL model consists of five convolutional layers and one fully-connected layer, with $V = 9.2\times 10^5$ total model parameters. For CIFAR-10, each local FL model is a standard ResNet-18 that consists of 17 convolutional layers and one fully-connected layer, with $V = 1.17 \times 10^7$ total model parameters.

For the MLP described in Sec.~\ref{ssec:mlp}, we focus on predicting the parameters of the last layer in each ML model (i.e., the fully connected layers). This leads to $V' = 640$ for both the CNN in Fashion-MNIST and the Resnet-18 in CIFAR-10. To restrict the ANN's total size to a small fraction of the ML model in each case, we set $D = 160$ for Fashion-MNIST and $D = 320$ for CIFAR-10, in line with CIFAR-10's higher complexity (leading to the MLP being 22\% and 3\% of the model sizes, respectively). As we will see in Sec.~\ref{ssec:robust}, we find that this retains a strong MSE performance for each dataset. Similar considerations can be made for other ML tasks.

Throughout the simulations, we will consider data samples to be allocated across devices in either an independent and identically distributed (IID) or non-IID manner. In the IID data setting, each device is allocated datapoints from all 10 class labels, while in the non-IID case, they each only receive datapoints from a fraction of the labels, assumed to be 6 by default \cite{YCHX,hosseinalipour2023parallel}.

It is worth noting that the convolution kernels exhibit sensitivity to errors, which significantly degrade the performance of the FL model, as will be demonstrated in the simulations. However, considering the relatively small number of parameters in convolution kernels, it is viable to utilize traditional methods such as orthogonal frequency division multiple access for transmitting the kernel parameters.

Finally, in some of our experiments (Fig.~\ref{Convergencetime}, Table~\ref{tab:latency_FMNIST}, Table~\ref{tab:latency_CIFAR}, Fig.~\ref{TrainingTime}), we will conduct accuracy comparisons over time/latency incurred rather than communication rounds. These calculations employ standard models for communication and computation delays~\cite{ZZY,DTse1}, which we present in Appendix~\ref{app:latency}. For experiments that refer to convergence (in particular, Tables~\ref{tab:latency_FMNIST} \&~\ref{tab:latency_CIFAR}), algorithms are considered to have converged when the value of the FL loss variance calculated over five consecutive iterations is less than $0.001$.

% In particular, each device has 2000 data samples for training a fully-connected neural network (FNN) that consists of one full-connection layer with 7840 (=28$\times$28$\times$10) model parameters on MNIST dataset and training a FNN that consists of three full-connection layers with 83900 (= 28$\times$28$\times$100 + 100$\times$50 + 50$\times$10) model parameters on Fashion-MNIST dataset. 
% The PS uses one MLP that consists of three layers to predict the FL gradient vector of each user. We assume that all local datasets are independent and identically distributed across the devices. All FL algorithms are considered to be converged when the value of the FL loss variance calculated over 20 consecutive iterations is less than 0.001.

\begin{table}
\centering
\vspace{-0.3cm}
\renewcommand\arraystretch{1}
\caption{Default Simulation Parameters} 
\vspace{-0.2cm}
\setlength{\tabcolsep}{1mm}{
  \begin{tabular}{|c|c||c|c||c|c|}
  \hline
\textbf{Parameters}&\textbf{Values}&\textbf{Parameters}&\textbf{Values}&\textbf{Parameters}&\textbf{Values}\\
  \hline
  \emph{K}& 20 &\emph{M}& 64 & $\sigma^2$& -90 dBW\\
  \hline
  $N_r$& 2 & $N_t$ & 2  &$N_k$& 2000 \\
  \hline
  $P_0$& 1 mW & $T$ & 50 &$W$& 7840\\
  \hline
  $N_I$ & 28 & $N_O$	& 10 & $\lambda $& 0.01 \\
  \hline
  $r$ & 1500 m & $D$	& 160 & $L_0 $& 100 \\
  \hline
 \end{tabular}}
 \vspace{-0.2cm}
\end{table}

\subsection{Training Speed and Accuracy Comparisons}

\begin{figure}
\centering
\includegraphics[width=9cm]{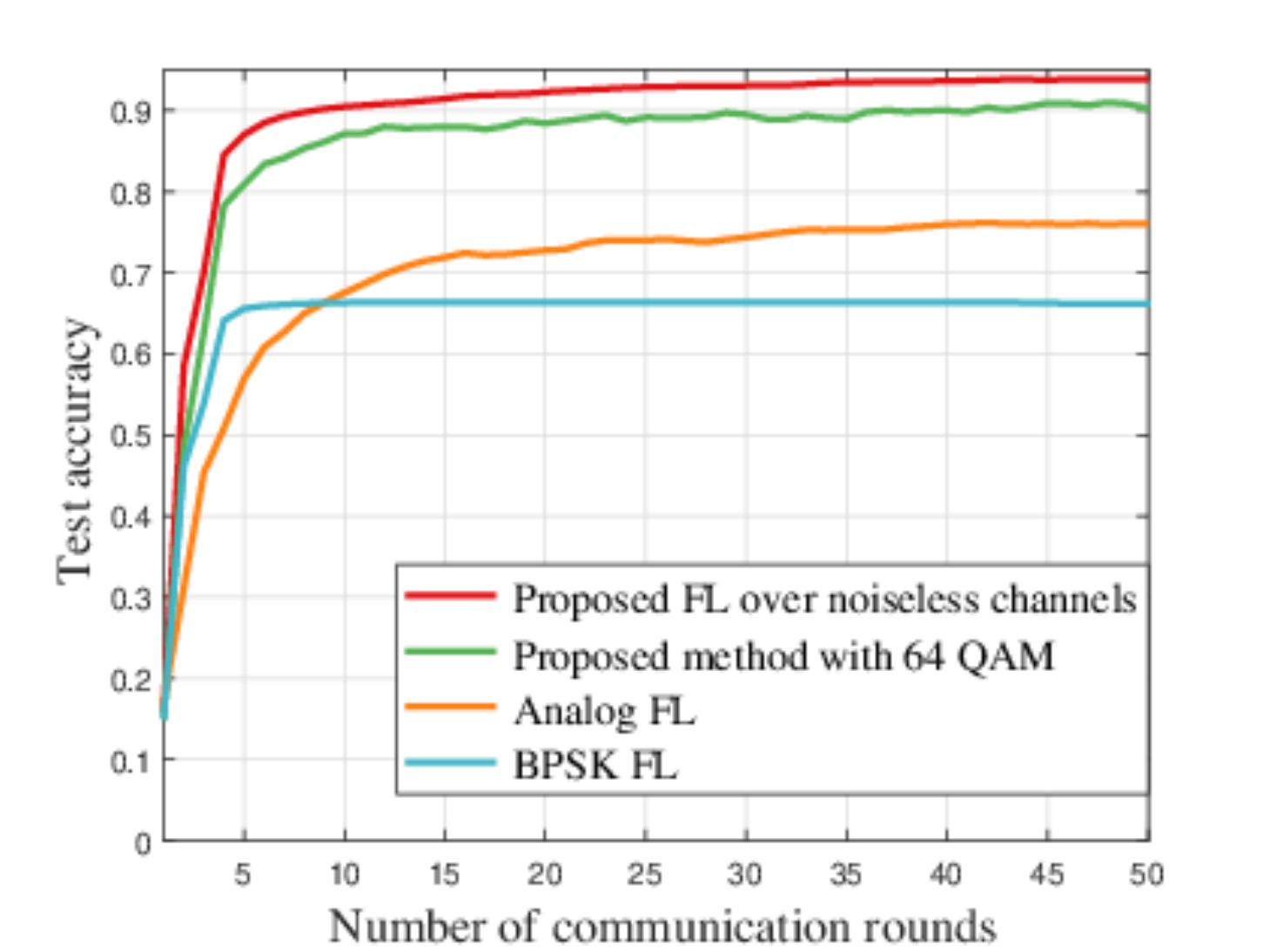}
\centering
\vspace{-0.4cm}
\caption{Testing accuracy of the proposed AirComp-based FL system over communication rounds, for an IID data allocation on the Fashion-MNIST task. We see that our proposed method provides substantial improvements over the baselines, approaching the noiseless channel case.}
\vspace{-0.3cm}
\label{SNRCNN1}
\end{figure}

We first conduct several experiments comparing our methodology to the baselines in terms of convergence speed and accuracy. In Fig. \ref{SNRCNN1}, we show how the test accuracy of all considered algorithms changes over communication rounds, on the Fashion-MNIST task, for the IID data setting. In this figure, we can see that, the proposed algorithm improves the test accuracy by up to 15.5\% and 24.5\% compared to analog FL and BPSK FL, respectively. From Fig. \ref{SNRCNN1}, we can also see that the performance of analog FL experiences noticeable fluctuations. This is due to the noise over wireless channels introducing dynamic errors into FL parameter transmission process, thus affecting FL test accuracy. We also see that the proposed method without using MLP for predicting FL gradients cannot converge, which verifies that the optimal beamforming matrices design depends on the prediction of FL gradients.

\begin{figure}
\centering
\includegraphics[width=9cm]{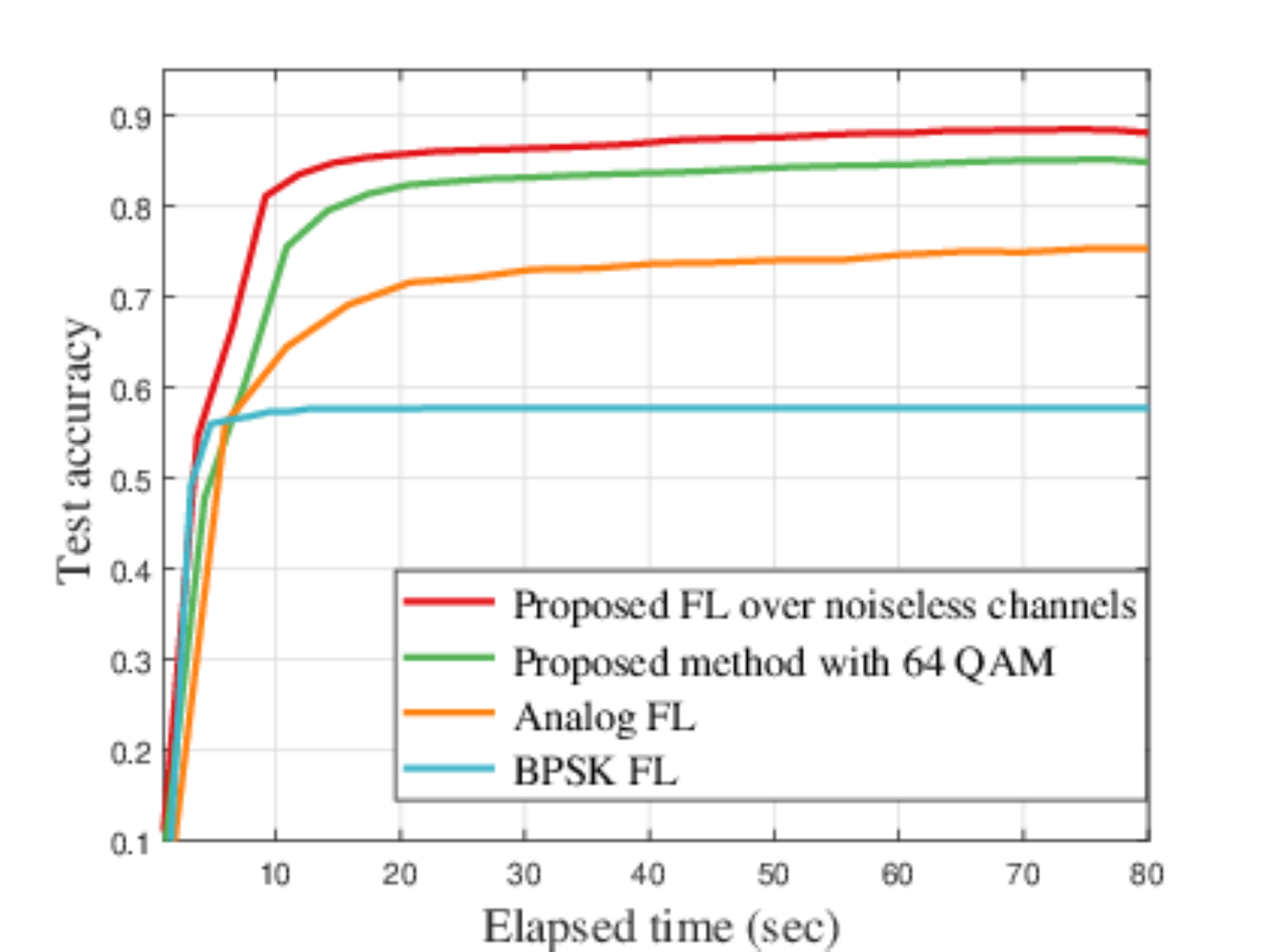}
\centering
\vspace{-0.4cm}
\caption{Test accuracy of the proposed AirComp-based FL system over time on the Fashion-MNIST task, for an non-IID data allocation. The overall trends are similar to those observed in Fig. \ref{SNRCNN1}.}
\vspace{-0.5cm}
\label{Convergencetime}
\end{figure}

Fig. \ref{Convergencetime} shows how the test accuracy changes over time elapsed, measured in seconds, as opposed to iterations. Here, the elapsed time consists of the FL model training time, inference time that is used for predicting FL parameters, and the model transmission time. The overall trends are similar to those observed in Fig. \ref{SNRCNN1}: compared to BPSK FL, while our proposed methodology converges more slowly, it ends up improving the test accuracy by up to 27\%. This is because the proposed FL uses more bits instead of one bit in BPSK FL to represent each FL parameter, thus increasing the dynamics of global FL model generation. In addition, this figure also shows that the proposed algorithm can reduce the convergence time by 20.3\% compared to analog FL. This finding underscores the fact that although the utilization of MLP introduces additional inference latency, it ultimately results in a decrease in the overall convergence time of FL. We will further investigate the latency vs. accuracy tradeoff in Tables \ref{tab:latency_FMNIST} and \ref{tab:latency_CIFAR}. %In contras \ref{Convergencetime} also shows that the test accuracy of BPSK FL is $61\%$.

\begin{figure}[t]
\centering
\includegraphics[width=9cm]{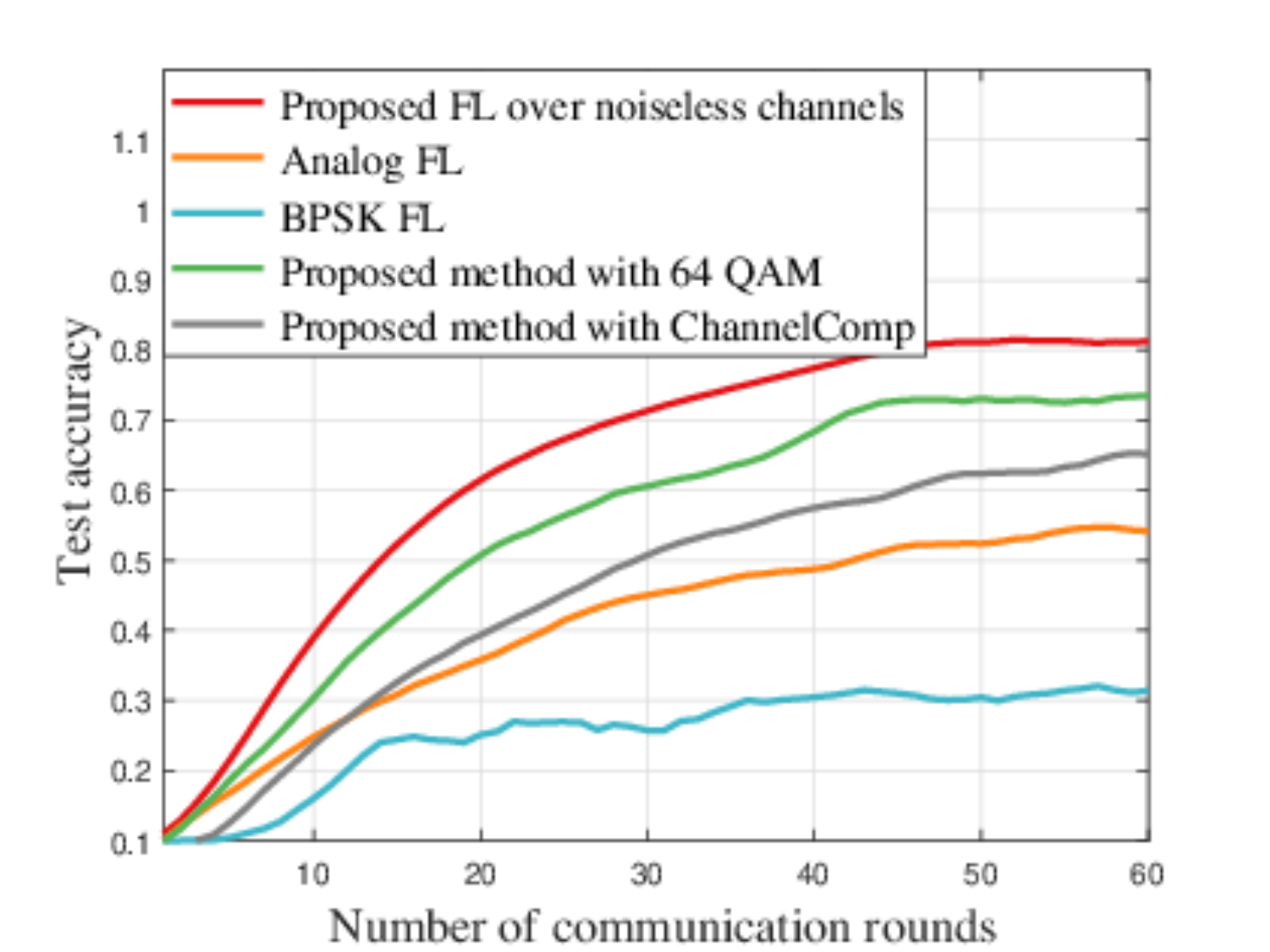}
\vspace{-0.3cm}
\caption{Test accuracy of the proposed AirComp-based
FL system across communication rounds, for an IID data allocation on the CIFAR-10 dataset. Compared to Fig. \ref{SNRCNN1}, we observe significant improvements over the baselines. The larger gap from the noiseless channel case shows the impact of noise on more complex learning tasks (CIFAR-10 vs. Fashion-MNIST).}
\vspace{-0.3cm}
\label{cifar10_iid}
\end{figure}

Fig. \ref{cifar10_iid} shows how the test accuracy changes as the number of communication rounds varies on the CIFAR-10 dataset, for the IID data distribution setting. Overall, we make similar observations from this figure as for Fashion-MNIST in Fig. \ref{SNRCNN1}. We see that our proposed method can improve the test accuracy by up to 20\% compared to analog FL, due to the proposed FL approach utilizing digital modulation (i.e., 64 QAM) to mitigate channel impairments and misalignments. We can also see that BPSK FL can only achieve 30\% test accuracy. The lower overall accuracies of each algorithm are consistent with CIFAR-10 being a more difficult learning task than Fashion-MNIST. Along these lines, the impact of fading and additive white Gaussian noise is stronger for this case compared with the Fashion-MNIST experiments, which can be explained by CIFAR-10 having an ML model that is more complex, causing the results to be more sensitive to impairments. Fig. \ref{cifar10_iid} also shows that our methodology with the ANN model prediction strategy leads to an improvement of up to 10\% compared to employing ChannelComp \cite{SJC}. This confirms that our prediction-based strategy obtains smaller errors compared to the transmission error minimization employed by this competing approach.
%Fig. 6 of \cite{SJC} contains further analysis of ChannelComp's error.}

%with known CSI, our ANN model prediction strategy leads to an improvement of up to 10\% compared to ChannelComp. This is because ChannelComp proposed in \cite{SJC} minimizes the difference between the demodulated values and the true values by optimizing a tabular function. However, this difference cannot be completely eliminated, even when the CSI is known, as shown in Fig. 6 of \cite{SJC}. In contrast, our proposed algorithm exhibits smaller prediction errors compared to this optimized error. Therefore, our proposed algorithm outperforms ChannelComp in terms of test accuracy of FL.}% thus resulting in a degeneration of test accuracy.}

\begin{figure}[t]
\centering
\includegraphics[width=9cm]{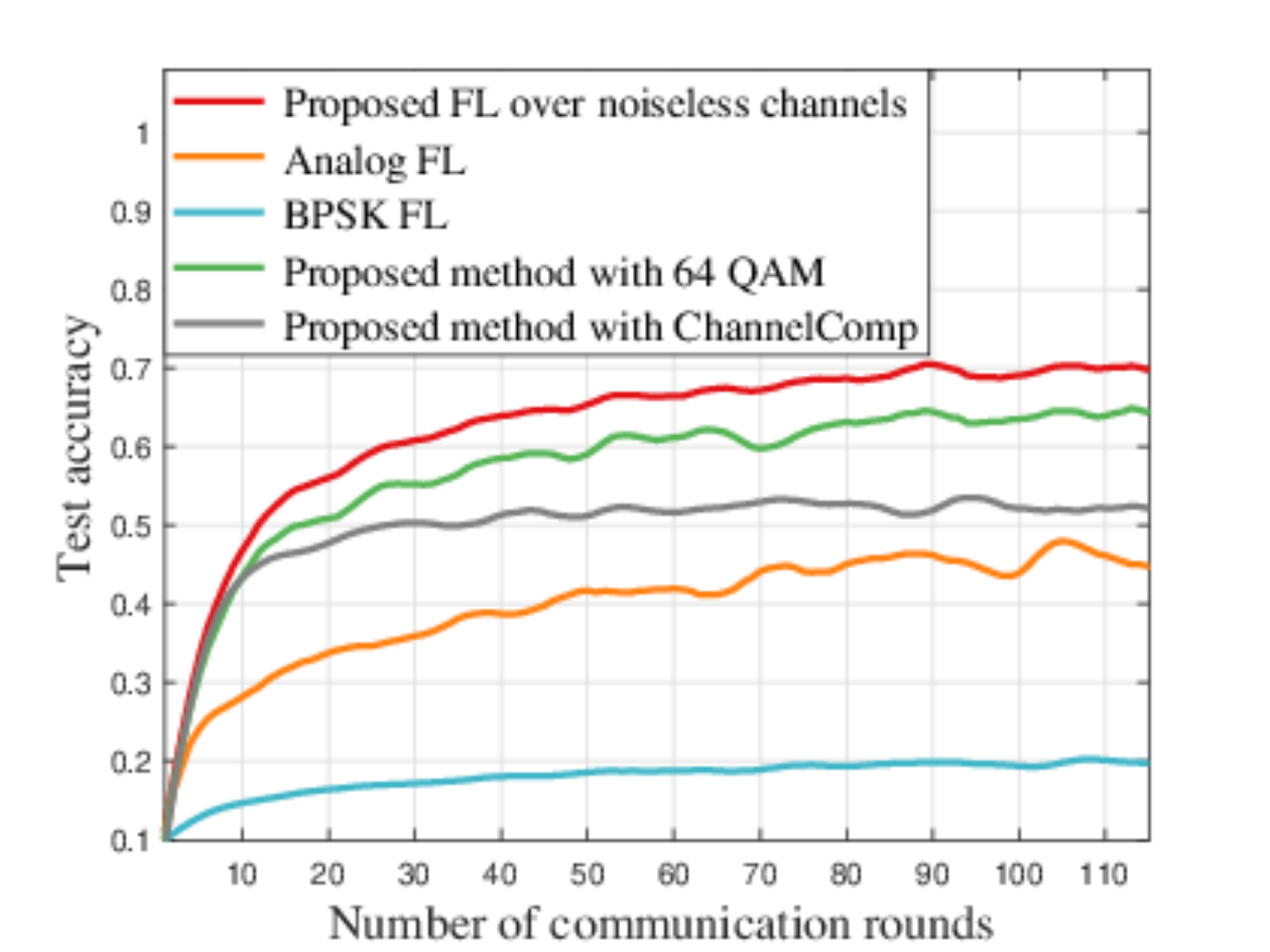}
\vspace{-0.3cm}
\caption{Test accuracy of the proposed AirComp-based
FL system across communication rounds, for a non-IID data allocation on the CIFAR-10 task. The overall trends are consistent with the IID case in Fig. \ref{cifar10_iid}, showing that our methodology obtains improvements under different data partitions.}
\vspace{-0.3cm}
\label{CIFRA_noniid}
\end{figure}

In Fig. \ref{CIFRA_noniid}, we repeat the experiment from Fig. \ref{cifar10_iid}, but this time under a non-IID setting, where the data samples of each device are drawn from 6 (instead of 10) classes. From this figure, we see that the proposed FL with 64-QAM can improve the test accuracy by up to 18\% and 13\% compared with analog FL and ChannelComp, similar to in the IID case. Fig. \ref{CIFRA_noniid} also shows that under noiseless channels, the accuracy of the AirComp-based system is 6\% better, consistent with the IID case.

Finally, in Tables \ref{tab:latency_FMNIST} and \ref{tab:latency_CIFAR}, we compare the performance of the proposed method with that of analog FL and BPSK FL in terms of (i) latency per iteration, (ii) overall training time, (iii) average rounds to convergence, and (iv) converged test accuracy, on the Fashion-MNIST and CIFAR-10 datasets, respectively, for the non-IID cases. Compared to analog FL we see that while our method incurs a higher latency in each round (e.g., $1.26$ vs. $1.03$ sec in Fashion-MNIST), it has a smaller latency across the entire training process (e.g., $18.90$ vs. $23.69$ sec in Fashion-MNIST), since it requires less total rounds to converge. Employing the MLP within our method introduces additional latency and computational overhead, but aids in designing the beamforming matrices to reduce transmission errors, thereby reducing the total number of rounds required. BPSK FL has the lowest latencies of all, but suffers from poor testing accuracy at convergence, due to the fact that it quantizes the FL parameters to a single bit. We observe too that the per-round and total latencies are higher in CIFAR-10, which is the most complex of the tasks, with the largest ML model being trained and transmitted.

\begin{table}[h]
\caption{Latency vs. Accuracy Tradeoff for Training on the Fashion-MNIST Dataset.}
\vspace{-0.3cm}
\centering
\def\temptablewidth{0.5\textwidth}
{\rule{\temptablewidth}{1pt}}
\scriptsize
%\footnotesize
\begin{tabular*}{\temptablewidth}{@{\extracolsep{\fill}}|c|c|c|c|c|}
&\makecell{Avg. Lat. for\\  Each Round} & \makecell{Avg. Lat. for\\ Entire Training}&  \makecell{Avg. Rounds \\  to Converge} &  \makecell{Converged \\  Test Acc.}\\
\hline
Proposed & 1.26 s& 18.90 s& 15 & 90\%\\
\hline
Analog FL& 1.03 s& 23.69 s& 23 & 76\%\\
\hline
BPSK FL& 0.21 s& 1.05 s& 5 &67\%\\
\hline
\end{tabular*}
{\rule{\temptablewidth}{1pt}}
\label{tab:latency_FMNIST}
\end{table}
\vspace{-0.3cm}

\begin{table}[t]
\caption{Latency vs. Accuracy Tradeoff for Training on the CIFAR-10 Dataset.}
\vspace{-0.3cm}
\centering
\def\temptablewidth{0.5\textwidth}
{\rule{\temptablewidth}{1pt}}  %根据使用情况灵活设置，线的粗细
\scriptsize
\begin{tabular*}{\temptablewidth}{@{\extracolsep{\fill}}|c|c|c|c|c|}
&\makecell{Avg. Lat. for\\  Each Round} & \makecell{Avg. Lat. for\\ Entire Training}&  \makecell{Avg. Rounds \\  to Converge} &  \makecell{Converged \\  Test Acc.}\\
\hline
Proposed & 21.46 s & 966 s& 45 &73\%\\
\hline
Analog FL& 20.92 s& 1255 s& 60 &52\%\\
\hline
BPSK FL& 2.95 s& 118 s& 40 &31\%\\
\hline
\end{tabular*}
{\rule{\temptablewidth}{1pt}}
\label{tab:latency_CIFAR}
\end{table}

\subsection{Robustness of MLP Predictor}
\label{ssec:robust}
\begin{figure}[t]
\centering
\includegraphics[width=9cm]{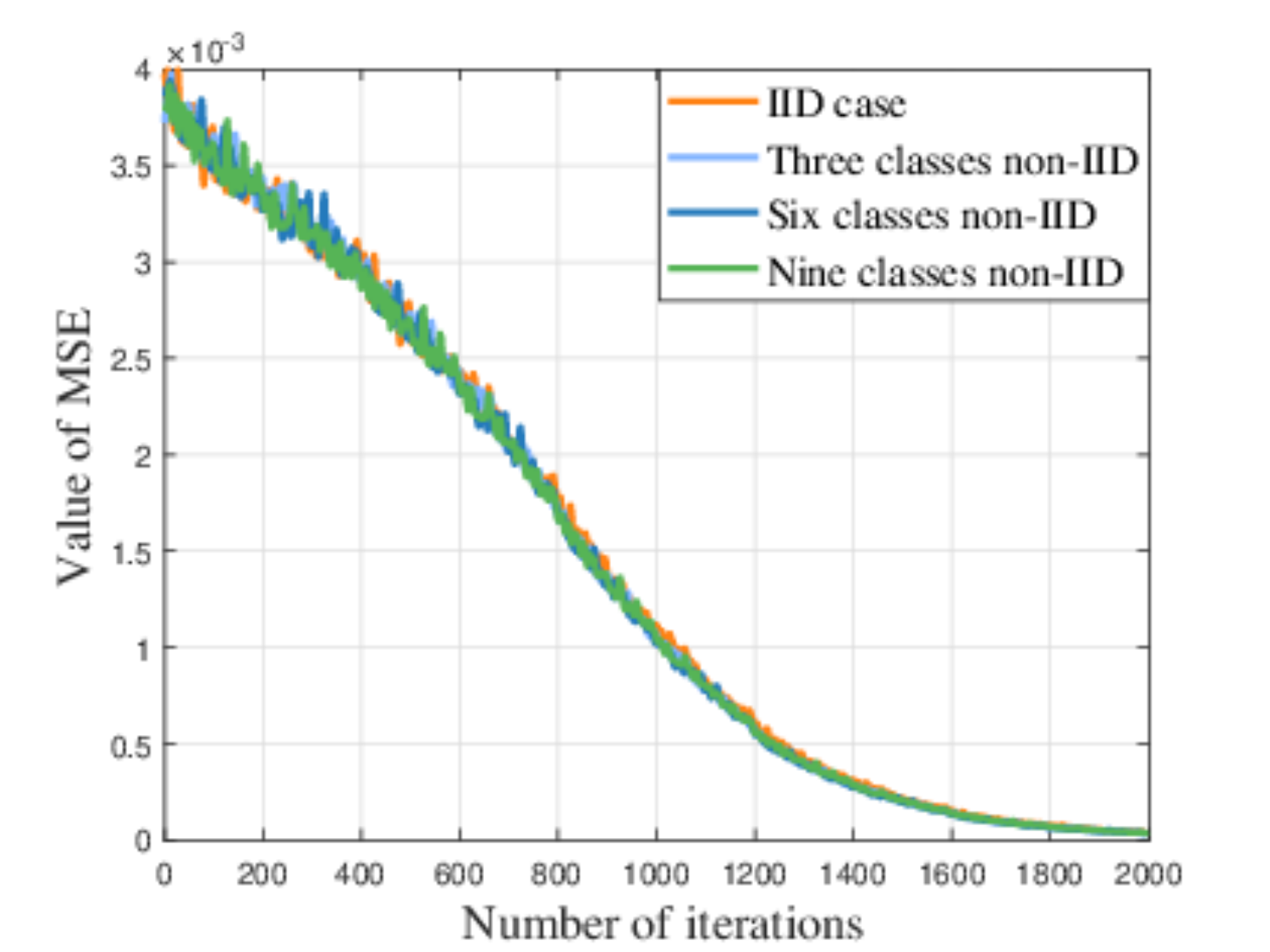}
\vspace{-0.3cm}
\caption{MSE performance of the proposed ANN-based model parameter predictor across training iterations on the CIFAR-10 dataset. We see that the ANN prediction quality is robust to the data distribution across devices.}
\vspace{-0.2cm}
\label{predict}
\end{figure}

Given the importance of the ANN-based predictor at the PS to our approach, we conduct further analysis to assess how it is impacted by the learning environment. First, in Fig. \ref{predict}, we show the performance of the MLP predicting FL parameters on the CIFAR-10 dataset as the data partitioning varies across devices. In the non-IID cases, the data samples of each device are from a fraction of the labels: 3 of 10, 6 of 10, and 9  of 10, respectively. The MSE is used to evaluate the performance of the MLP on the test dataset at each iteration of training. In this figure, we see that the proposed ANN algorithm achieves a predominantly linear convergence speed prior to tapering off. This is because the proposed ANN has only three fully-connected layers thus having a relatively low computational complexity. Fig. \ref{predict} also shows that as the proposed ANN method converges, the MSE approaches $0$, which implies the proposed ANN approach can predict FL parameters accurately. We also see that the values of the MLP MSE in both the IID and non-IID data distributions are similar. This indicates that MLP is capable of capturing the nonlinear relationships of the FL parameters for different distributions of data samples across the devices.

\begin{table*}[t]
\caption{\small MSE performance of the proposed ANN-based model parameter predictor when the device models are transmitted over different channel conditions, for CIFAR-10 with a non-IID data partition.}
\centering
\vspace{-0.3cm}
\small
\def\temptablewidth{1.0\textwidth}
{\rule{\temptablewidth}{1pt}} 
\begin{tabular*}{\temptablewidth}{@{\extracolsep{\fill}}|c|c|c|c|c|c|}
 &\makecell{ MLP with perfect\\ communication} & \makecell{Rayleigh, \\ $\sigma^2 = -90$ dBm}& \makecell{Rayleigh, \\ $\sigma^2 = -95$ dBm} & \makecell{Rician, \\ $\sigma^2 = -90$ dBm, \\ $k = 7$}  & \makecell{Rician, \\ $\sigma^2 = -90$ dBm, \\ $k = 10$}\\
\hline
\makecell{ Converged \\ MSE ($\times 10^{-7}$)} &7.73 & 13.21 & 11.14 & 8.51 & 7.80\\
\hline
\end{tabular*}
{\rule{\temptablewidth}{1pt}}
\label{tab:ANN-channel}
\end{table*}

Next, in Table \ref{tab:ANN-channel}, we show the MSE of the MLP when the FL models of devices are transmitted over different channel conditions, on the CIFAR-10 dataset. Here, we consider a non-IID data distribution where the data samples of each device are from 6 of 10 labels. We consider the implementation of the proposed framework over four channel conditions: (a) a Rayleigh model with noise variance $\sigma^2 = -90$ dBm, (b) a Rayleigh model with noise variance $\sigma^2 = -95$ dBm, (c) a Rician model with noise variance $\sigma^2 = -90$ dBm and Rician factor $k = 7$, and (d) a Rician model with noise variance $\sigma^2 = -90$ dBm and Rician factor $k = 10$. The pathloss exponent is set to $\beta = 2$ for Rayleigh and $\beta = 4$ for Rician. The MSE performance of the MLP without transmission errors is considered as an optimal baseline. Overall, we can see that the MLP demonstrates satisfactory performance in terms of MSE regardless of the channel model, which indicates the robustness of the proposed ANN algorithm. We also notice that the MSE of the MLP transmitted through the Rician fading channel is lower than the MSE of the MLP transmitted through the Rayleigh fading channel due to the influence of the line-of-sight signal. Additionally, as the Rician factor increases and the noise variance decreases, the MSE of the MLP transmitted over different channels approaches the optimal baseline.

\begin{figure}[t]
\centering
\includegraphics[width=9cm]{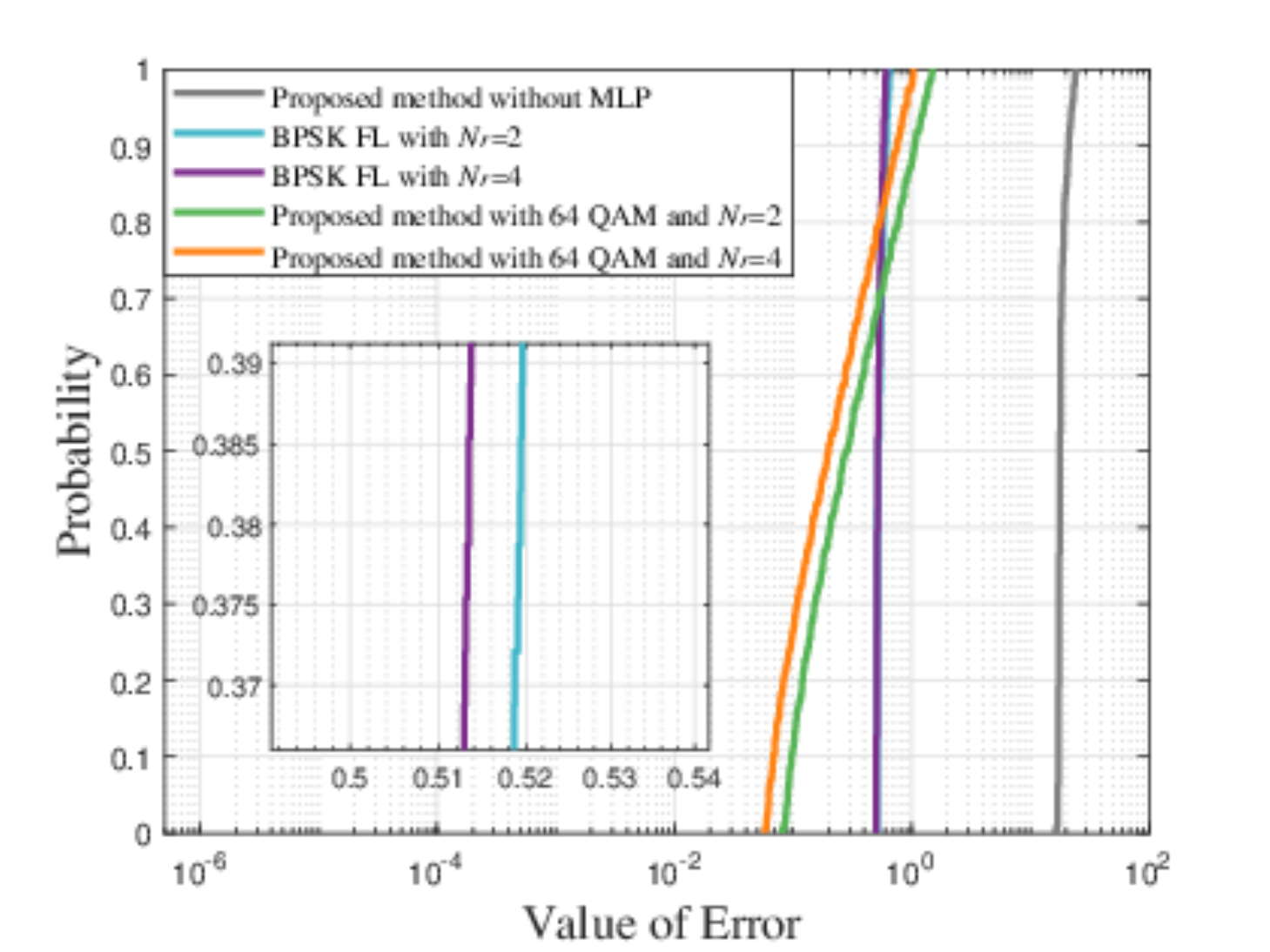}
\vspace{-0.4cm}
\caption{Cumulative distribution function (CDF) of the errors in the aggregated models, for training on Fashion-MNIST dataset under the IID data partition.}
\vspace{-0.2cm}
\label{CDF}
\end{figure}

Finally, Fig. \ref{CDF} assesses the error in the obtained aggregated model $\bm g_t$ at the PS for different methods. The error is defined as the sum of the distances between all weights in the aggregated model at the PS and that in the true average of the models across the devices. The results are shown as the cumulative distribution function (CDF) of the value of the errors obtained over training rounds. We can see that the proposed method achieves a lower error rate compared to the proposed FL without using MLP for FL gradient prediction. This is because without predicting FL gradient vectors, the PS cannot proactively adjust the transmit and receive beamforming matrices to minimize transmission errors and can only use fixed beamforming design that directly aggregates all local models via linear superimposition. This linear superimposition is not available for digital modulation schemes since digital modulation may introduce complex mapping relationships between bits and symbols thus resulting in additional demodulation errors. %{\color{blue}Fig. \ref{CDF} also shows that the performance for baseline c) with different receiver antennas are almost same. This is due to the fact that baseline c) adopts BPSK for digital modulation that minimizes the transmission error and hence, an increase of the number of  receiver antennas can not continue to reduce the transmission error. However, the error in Fig. \ref{CDF} consists of transmission error and quantization error that is not minimized in BPSK.} 

\subsection{Impacts of Channel Conditions and SNR}

\begin{figure}[t]
\centering
\includegraphics[width=9cm]{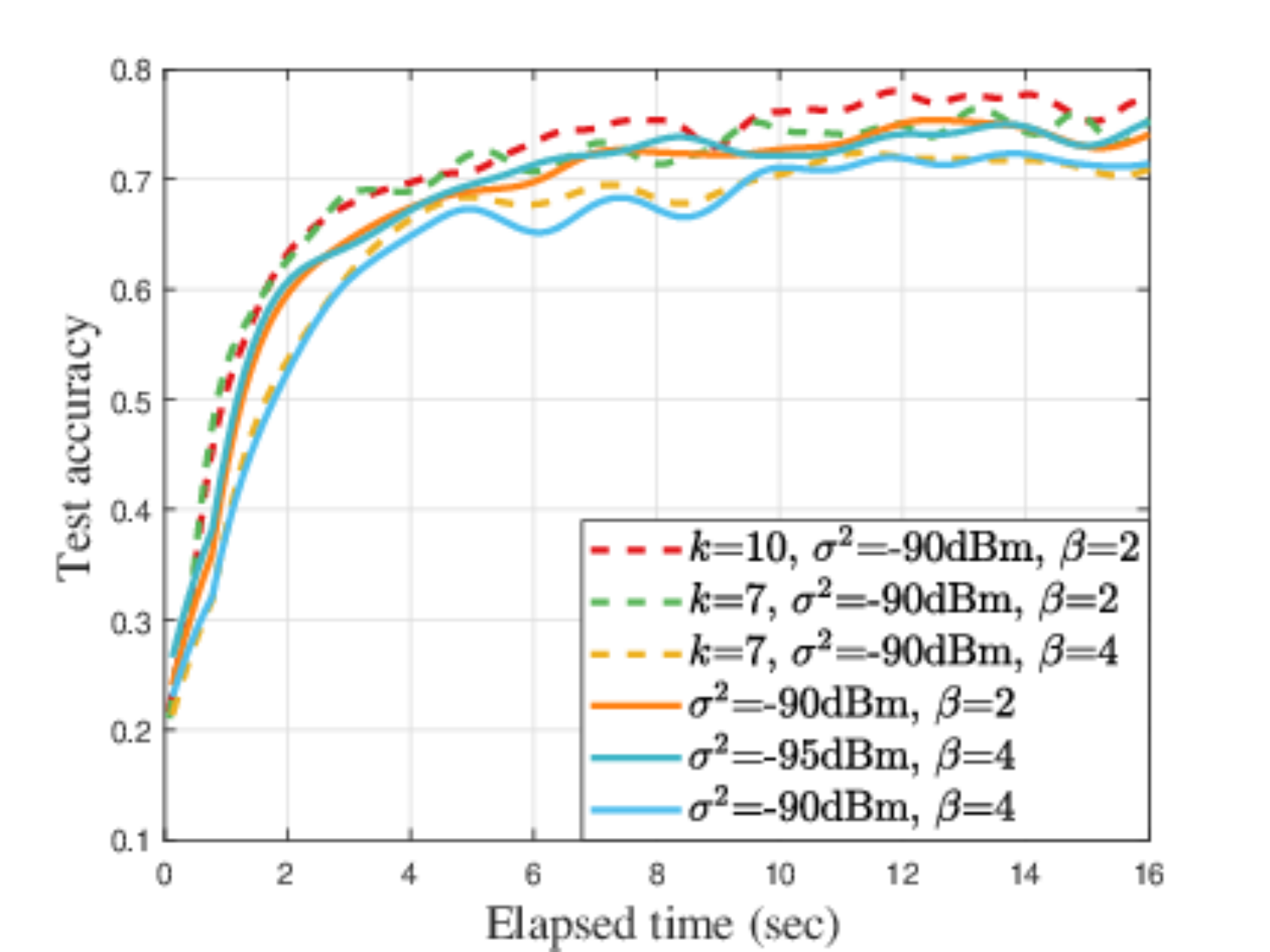}
\vspace{-0.4cm}
\caption{Impact of Rician (dashed lines) and Raleigh (solid lines) channel models on training performance, for the Fashion-MNIST dataset under an IID data partition. Overall, we see that our methodology is reasonably robust to variations in the channel model.}
\vspace{-0.2cm}
\label{TrainingTime}
\end{figure}

We next investigate the impact of channel conditions on the performance of our methodology. In Fig. \ref{TrainingTime}, we show the performance of the proposed method on the Fashion-MNIST dataset under various channel models. In particular, we consider the Rayleigh fading channel and the Rician fading channel. For each model, we consider three different settings, varying the noise variance $\sigma^2 = -90$ dBm, pathloss exponent $\beta$, and Rician factor $k$. Overall, from this figure, we can see that the proposed algorithm is reasonably robust to the specific channel conditions, with the testing accuracy staying within a window of $\pm 0.04$. The best performance occurs when the channel is Rician and $k = 10$. This is consistent with the Rician factor being the ratio of the power of the dominant path component to the power of the whole signal. Thus, $k = 10$ is the case of the largest dominant path, which reduces FL transmission errors and improves the FL performance.

\begin{figure}
\centering
\includegraphics[width=9cm]{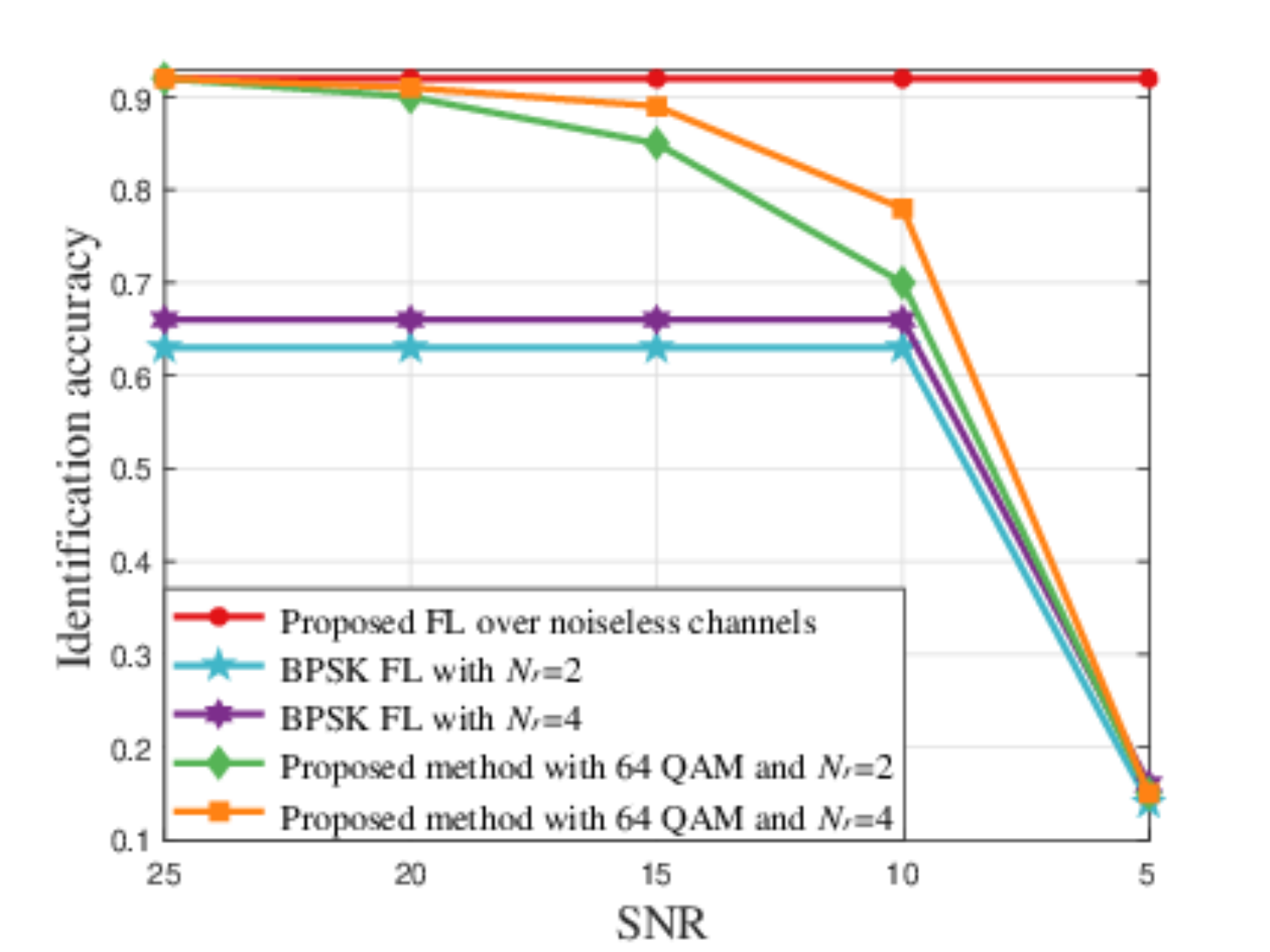}
\vspace{-0.4cm}
\caption{Impact of SNR on test accuracy for the Fashion-MNIST dataset, for an IID data partition. We see that the performance of all methods begins to decay once the SNR drops below 10 dB.}
\vspace{-0.2cm}
\label{SNR_ACC}
\end{figure}

We next consider the impact of the received SNR. Fig. \ref{SNR_ACC} shows how the test accuracy of considered FL algorithms changes as SNR decreases, for the Fashion-MNIST dataset. In Fig. \ref{SNR_ACC}, we can see that the test accuracy of the proposed method decreases as SNR decreases, while the test accuracy of BPSK FL remains unchanged, when the SNR is larger than 10 dB. Additionally, the test accuracy of BPSK FL is lower than the proposed method at any SNR values. This is because the quantization error in BPSK significantly affects the model training process and results in a degeneration of test accuracy. Fig. \ref{SNR_ACC} also shows that the proposed method with $N_r = 4$ receiver antennas can achieve 8\% gains in terms of test accuracy compared to that with $N_r = 2$ receiver antennas when SNR is 10 dB. Thus, an increase of the number of receiver antennas can improve the test accuracy in the proposed FL framework. This is because an increase of the number of receiver antennas enables the PS to exploit transmit diversity and reduce transmission error in the AirComp-based system.
%From Fig. \ref{SNR_ACC}, we can also see that as SNR is 5~dB, the test accuracy of the proposed algorithm and BPSK FL decreases to 0.1.

\begin{figure}
\centering
\includegraphics[width=9cm]{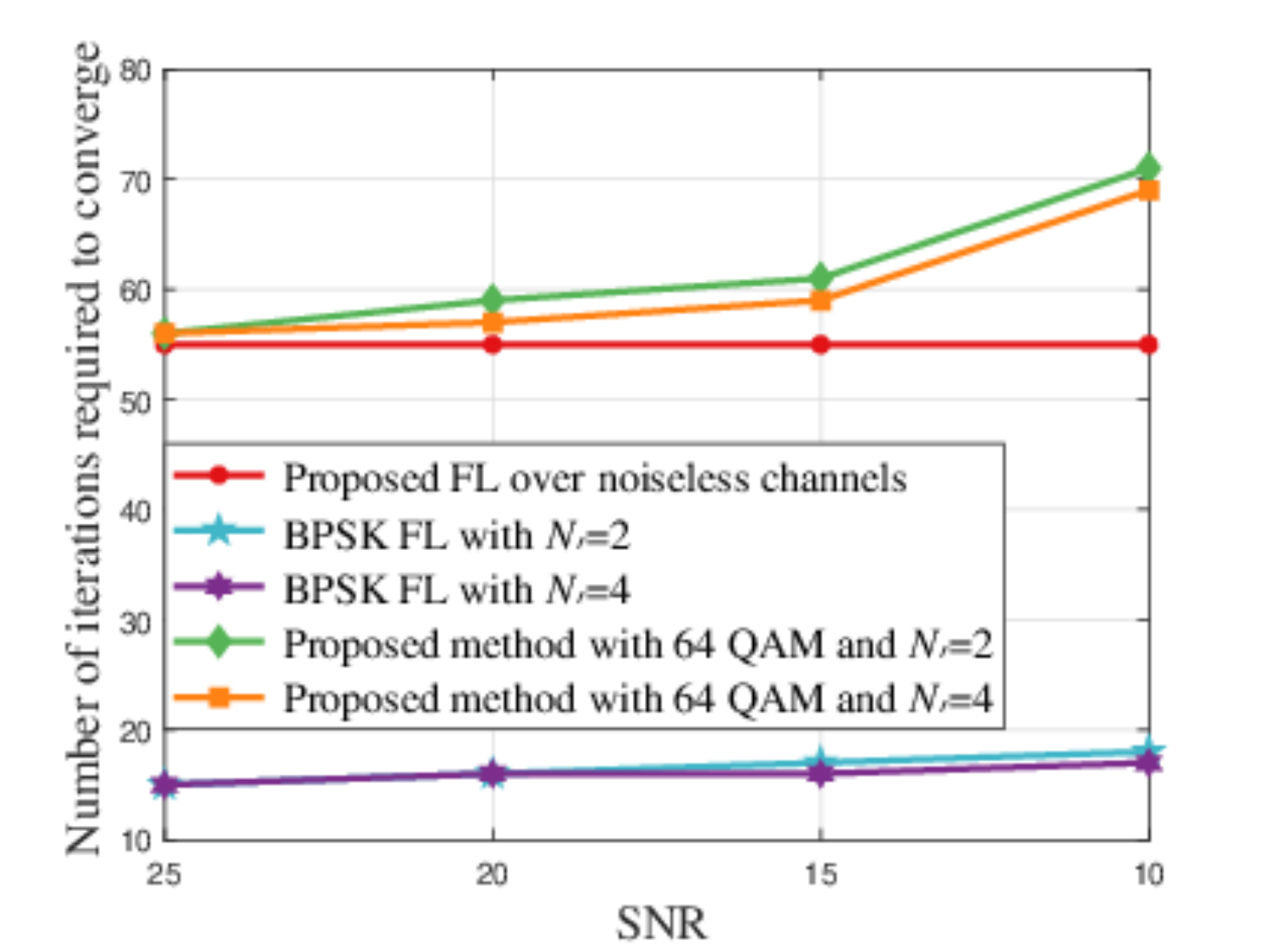}
\vspace{-0.5cm}
\caption{Impact of SNR on convergence performance for the Fashion-MNIST dataset, for the same settings as in Fig. \ref{SNR_ACC}. We see that our method converges significantly faster at higher SNRs, with less transmission errors occurring.}
\label{SNR_conv}
\vspace{-0.2cm}
\end{figure}

Finally, in Fig. \ref{SNR_conv}, we show how the number of communication rounds that the considered FL algorithms require to converge changes as SNR decreases, for the same settings as in Fig. \ref{SNR_ACC}. We can see that, compared with BPSK FL, the number of communication rounds required to converge for our methodology decreases noticeably as the SNR increases. This is due to the fact that, as SNR decreases, the probability of introducing additional transmission errors increases thus reducing the FL convergence speed. 

\subsection{Impact of Network Size}

\begin{figure}
\centering
\includegraphics[width=9cm]{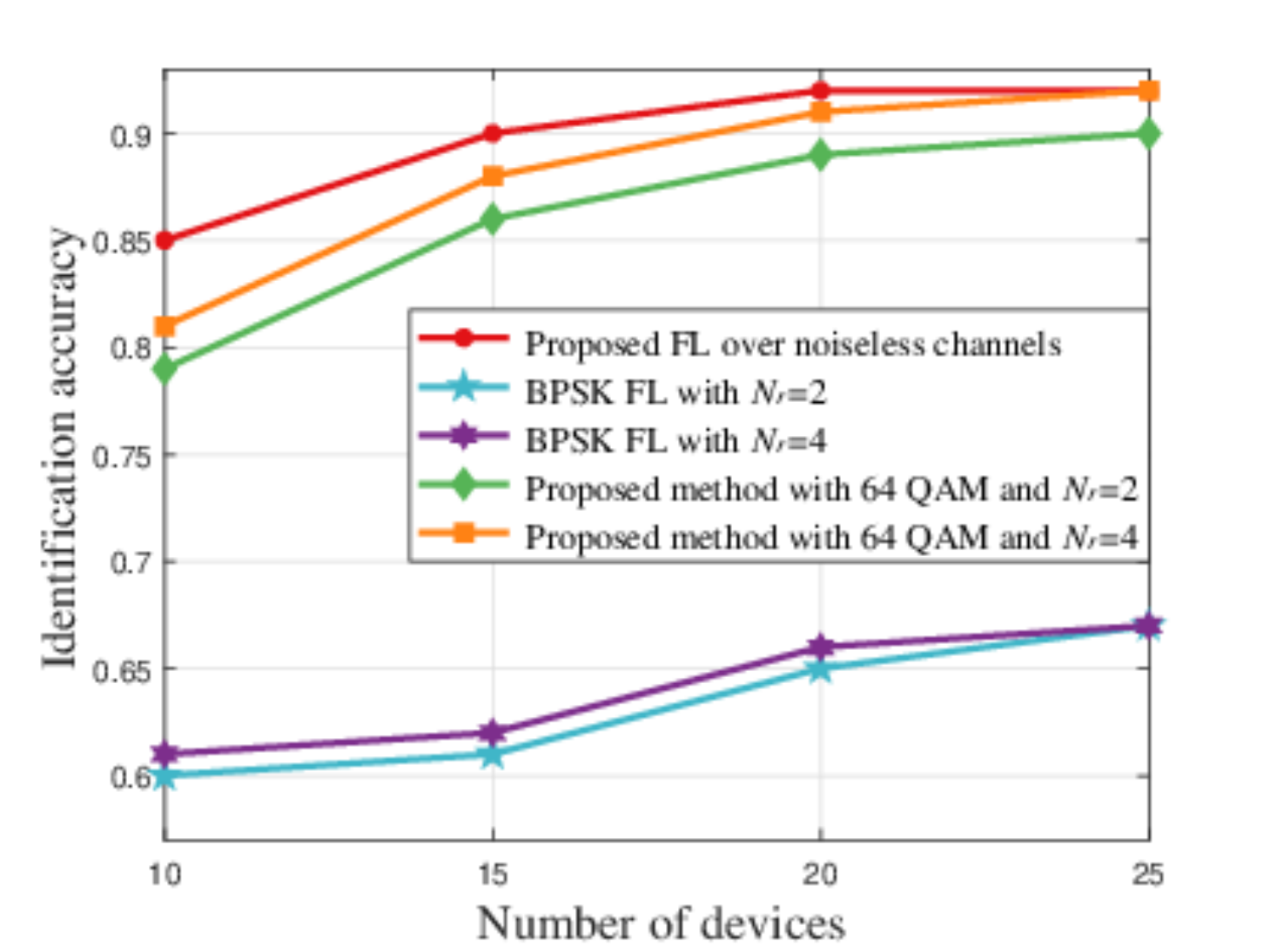}
\vspace{-0.4cm}
\caption{Impact of network size on obtained accuracy for the Fashion-MNIST dataset under an IID data partition. A larger network has a positive impact on accuracy performance.}
\vspace{-0.2cm}
\label{device_Acc}
\end{figure}

Finally, we consider the impact of the number of devices on the performance. Fig. \ref{device_Acc} shows how the test accuracy changes as the number of devices varies. In Fig. \ref{device_Acc}, we can see that the test accuracy increases as the number of devices increases. This is because as the number of devices increases, the number of data samples available in the network for training increases. This helps improve the performance both of our higher-order modulation method as well as for BPSK FL.
%Once the number of devices has reached 25, the testing accuracy performance has saturated. Also, from Fig. \ref{device_Acc}, we see that the proposed method can improve the test accuracy by 25\% compared to baseline BPSK FL. This is because the proposed FL framework enables the PS and the devices to utilize high-order digital modulation to reduce quantization errors. %while maintaining superimposition.
Fig. \ref{device_Acc} also shows that as the number of receiver antennas increases from $N_r = 2$ to $N_r = 4$, the test accuracy of the proposed methodology experiences a slight improvement, because of the PS's ability to exploit transmit diversity.
%This implies that an increase of the number of receiver antennas may not affect the gain for test accuracy under ideal channel conditions.

\begin{figure}
\centering
\includegraphics[width=9cm]{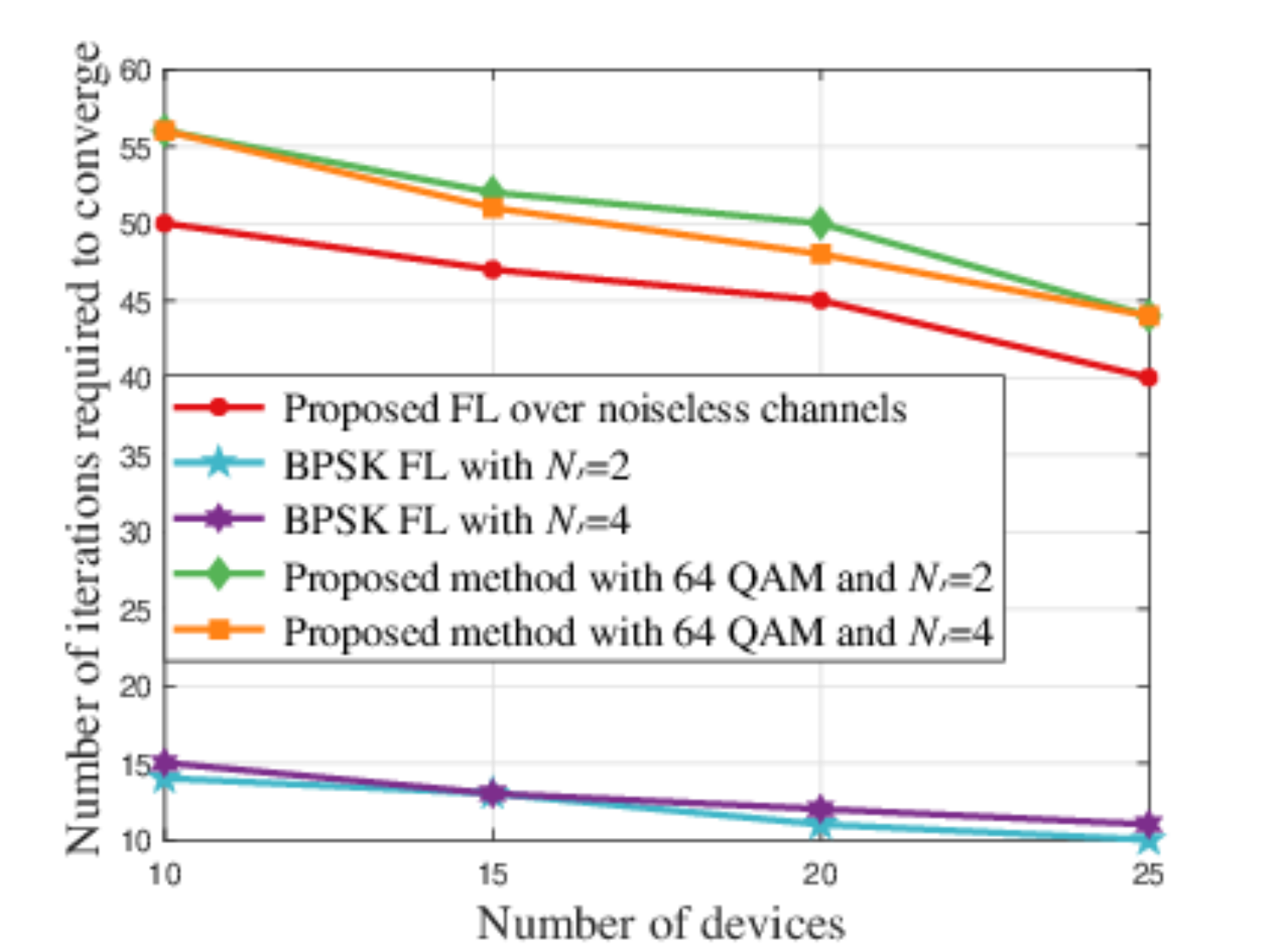}
\vspace{-0.3cm}
\caption{Impact of network size on convergence performance, for the same settings as Fig. \ref{device_Acc}. A larger network reduces the iterations required.}
\vspace{-0.2cm}
\label{device_Conv}
\end{figure}

In Fig. \ref{device_Conv}, we show how the number of rounds required to converge varies as the number of devices changes, for the same dataset and settings as Fig. \ref{device_Acc}. Overall, we see that as the number of devices increases, the number of rounds needed to converge decreases, in addition to improving the overall testing accuracy. Fig. \ref{device_Conv} also shows that our method can exploit a larger number of antennas to reduce the iterations required for convergence, with appropriate design of the beamforming matrices.
%the proposed method can achieve the same performance in terms of convergence speed compared to the proposed FL over noiseless channels, which illustrates that our proposed method can approach perfect model aggregation when considering fading and additive white Gaussian noise.

\section{Conclusion}

In this article, we have developed a novel framework that enables the implementation of FL algorithms over a digital MIMO and AirComp based system. We have formulated an optimization problem that jointly considers transmit and receive beamforming matrices for the minimization of FL training loss. To solve this problem, we analyzed the expected improvement of FL training loss between two adjacent iterations that depends on the digital modulation mode, the number of devices, and the design of beamforming matrices. To obtain this bound in practice, we introduced an ANN based algorithm to estimate the local FL models of all devices; then, the optimal solution of beamforming matrices can be determined based on the predicted FL model and the derived expected improvement of FL training loss. Numerical evaluation on real-world machine learning tasks demonstrated that the proposed methodology yields significant gains in classification accuracy and convergence speed compared to conventional approaches.

\appendices

\section{Proof of Theorem 1}
\label{app:proof}
To prove Theorem 1, we first rewrite $F(\bm g_{t+1})$ using the second-order Taylor
expansion and the property of the $L$-smooth in Assumption a), which can be expressed as
\begin{equation}\label{eq:A1}
	\begin{aligned}
		\!F\left({\bm g}_{t+1}\right) \leq F\left({\bm g}_{t}\right)+\left(  {\bm g}_{t+1}- {\bm g}_{t} \right)\nabla F\left({\bm g}_{t}\right)+\frac{L}{2}\left\| {\bm g}_{t+1}- {\bm g}_{t}\right\|^2.
	\end{aligned} \notag
\end{equation}
Let ${\bm g}_{t+1}-{\bm g}_{t}=\nabla F\left({\bm g}_{t}\right)-\bm o_t$ and the learning rate $\lambda= \frac{1}{L}$, we have
\begin{equation}\label{eq:A2}
\begin{split}
&\mathbb{E}\left(F\left({\bm g}_{t+1}\right)\right)-\mathbb{E}\left(F\left({\bm g}_{t}\right)\right) \\
\leq &-\lambda \mathbb{E}\left(\nabla F\left({\bm g}_{t}\right)-\bm o_t\right)\nabla F\left({\bm g}_{t}\right)+\frac{L\lambda^2}{2}\mathbb{E}\left\|\nabla F\left({\bm g}_{t}\right)-\bm o_t\right\|^2\\
\overset{(a)}{=} &-\frac{1}{2L}\mathbb{E}\left\|\nabla F\left({\bm g}_{t}\right)\right\|^2+\frac{1}{2L}\mathbb{E}\left(\left\| \bm o_t\right\|^2\right), \notag
\end{split}
\end{equation}
where (a) stems from the fact that $\frac{L\lambda^2}{2}\left\|\nabla F\left({\bm g}_{t}\right)-\bm o_t\right\|^2=\frac{1}{2L}\left\|\nabla F\left({\bm g}_{t}\right)\right\|^2-\frac{1}{L} {\bm o^T}\nabla F\left({\bm g}_{t}\right)+\frac{1}{2L}\mathbb{E}\left(\left\| \bm o_t\right\|^2\right)$ with $\bm o_t$ being a gradient deviation caused by the errors in local FL model transmission, which can be given as follows
\begin{align}\label{eq:A2}
&\mathbb{E}\left(\left\| \bm o_t\right\|^2\right) \notag\\
=&\mathbb{E}\left(\left\|\nabla F\left({\bm g}_{t}\right) - \left( {\bm g}_{t+1}-{\bm g}_{t} \right) \right\|^2\right) \notag \\
=&\mathbb{E}\left( \left\| \frac{\sum\limits_{k = 1}^{K}\sum\limits_{n=1}^{N_{k}}\nabla f\left(\bm g_t,\bm x_{n,k},\bm y_{n,k}\right)}{N} \right. \right. \notag \\ 
&\qquad \quad \left. \left. - l^{-1} \left( \bm {\hat s}_{t} \left(\sum\limits_{n\in \mathcal{N}_{k,t}}\nabla f\left(\bm g_t,\bm x_{n,k},\bm y_{n,k}\right) \right) \right) \right\|^2 \right) \notag\\
\leqslant & \mathbb{E}\!\left( \left\| \frac{\sum\limits_{k = 1}^{K}\sum\limits_{n=1}^{N_{k}}\!\!\nabla \! f\!\left(\bm g_t,\bm x_{n,k},\bm y_{n,k}\right)}{N} - \frac{\sum\limits_{k = 1}^{K}\sum\limits_{n=1}^{N_{k,t}}\!\!\nabla \!  f\!\left(\bm g_t,\bm x_{n,k},\bm y_{n,k}\right)}{\sum\limits_{k = 1}^{K} |\mathcal{N}_{k,t}|}  \right\| \right. \notag \\ 
& \quad \quad + \left\| \frac{\sum\limits_{k = 1}^{K}\sum\limits_{n=1}^{N_{k,t}}\nabla f\left(\bm g_t,\bm x_{n,k},\bm y_{n,k}\right)}{N} \right.  \notag \\ 
&\left. \left. \qquad \quad -l^{-1} \left( \frac{\sum\limits_{k = 1}^{K} l \left( \sum\limits_{n\in \mathcal{N}_{k,t}}\nabla f\left(\bm g_t,\bm x_{n,k},\bm y_{n,k}\right) \right)}{\sum\limits_{k = 1}^{K} |\mathcal{N}_{k,t}|} \right) \right\| \right. \notag\\
+ & \left. \left\| l^{-1} \left( \frac{\sum\limits_{k = 1}^{K} l \left( \sum\limits_{n\in \mathcal{N}_{k,t}}\nabla f\left(\bm g_t,\bm x_{n,k},\bm y_{n,k}\right)\! \right)}{\sum\limits_{k = 1}^{K} |\mathcal{N}_{k,t}|} \right)\right. \right. \notag \\ 
&\qquad \quad \left. \left.-l^{-1} \left( \bm {\hat s}_{t} \left(\sum\limits_{n\in \mathcal{N}_{k,t}}\nabla f\left(\bm g_t,\bm x_{n,k},\bm y_{n,k}\right) \right) \right) \right\|^2 \right) \notag \\
\end{align}
where $\nabla f\left({\bm g}_{t},\bm x_{n,k},\bm y_{n,k}\right)$ is the gradient trained by $\left(\bm x_{n,k},\bm y_{n,k}\right)$. $\bm {\hat s}_{t} \left(\sum\limits_{n\in \mathcal{N}_{k,t}}\nabla f\left(\bm g_t,\bm x_{n,k},\bm y_{n,k}\right) \right)=\frac{\bm B \sum\limits_{k = 1}^{K} \bm H_k \bm A_k l \left( \sum\limits_{n\in \mathcal{N}_{k,t}}\nabla f\left(\bm g_t, \bm x_{n,k}, \bm y_{n,k}\right)\right) +\bm B \bm n_t}{\sum\limits_{k = 1}^{K} |\mathcal{N}_{k,t}|}$ is the received signal. $\frac{\sum\limits_{k = 1}^{K} l \left( \sum\limits_{n\in \mathcal{N}_{k,t}}\nabla f\left(\bm g_t, \bm x_{n,k}, \bm y_{n,k}\right) \right)}{\sum\limits_{k = 1}^{K} |\mathcal{N}_{k,t}|}$ is the theoretical signal that is obtained via modulation at devices and demodulation at the PS without channel impairments and misalignments. Given $\nabla f\left({\bm g}_{t},\bm x_{n,k},\bm y_{n,k}\right)$, the first term in (23) represents the discrepancy of gradients introduced by random sampling of mini-batches, which can be rewritten as
\begin{equation}\label{eq:update33}
\begin{aligned}
&\mathbb{E} \left\|\frac{\sum\limits_{k = 1}^{K}\sum\limits_{n=1}^{N_{k}}\!\!\nabla \! f\left(\bm g_t,\bm x_{n,k},\bm y_{n,k}\right)}{N} - \frac{\sum\limits_{k = 1}^{K}\sum\limits_{n=1}^{N_{k,t}}\!\!\nabla \!  f\left(\bm g_t,\bm x_{n,k},\bm y_{n,k}\right)}{\sum\limits_{k = 1}^{K} |\mathcal{N}_{k,t}|} \right\| \notag \\
=&\mathbb{E}\left\| \sum\limits_{k = 1}^{K} \widehat{\bm w}_{k,t}- \sum\limits_{k = 1}^{K} \bm w_{k,t} \right\| \\
=&\sum\limits_{k = 1}^{K} \frac{\sigma^2_k}{|\mathcal{N}_{k}|},
\end{aligned}
\end{equation}
where $\widehat{\bm w}_{k,t}$ is the optimal updated local model without variance. Given $\frac{\sigma^2_k}{|\mathcal{N}_{k}|}$ on each device $k$, the variance bound of all devices is $\Upsilon= \sum\limits_{k = 1}^{K}  \frac{\sigma^2_k}{|\mathcal{N}_{k}|}$.

And then, we have $\mathbb{E}\left(\left\| \bm o_t\right\|^2\right)=\mathbb{E}\left(\left\| \bm e_t + \hat {\bm e_t}(\bm A_{t},\bm B_{t})\right\|^2\right)$ where
\begin{equation}\label{eq:A000}
\begin{split}
\bm e_t=&\left\| \frac{\sum\limits_{k = 1}^{K}  \sum\limits_{n\in \mathcal{N}_{k,t}} \nabla f\left(\bm g_t,\bm x_{n,k}, \bm y_{n,k}\right) }{\sum\limits_{k = 1}^{K} |\mathcal{N}_{k,t}|} \right. \\
&\left. -l^{-1}\left(\frac{\sum\limits_{k = 1}^{K} l \left( \sum\limits_{n\in \mathcal{N}_{k,t}}  \nabla f\left(\bm g_t,\bm x_{n,k}, \bm y_{n,k}\right) \right)}{\sum\limits_{k = 1}^{K} |\mathcal{N}_{k,t}|}\right)\right\| + \Upsilon
\end{split}
\end{equation}
and
\begin{equation}\label{eq:A002}
\begin{split}
&\hat {\bm e_t}(\bm A_{t},\bm B_{t})= l^{-1}\left(\frac{\sum\limits_{k = 1}^{K} l \left( \sum\limits_{n\in \mathcal{N}_{k,t}}  \nabla f\left(\bm g_t,\bm x_{n,k}, \bm y_{n,k}\right) \right)}{\sum\limits_{k = 1}^{K} |\mathcal{N}_{k,t}|}\right)\\
&-l^{-1} \left( \!\frac{\bm B_t \left(\sum\limits_{k = 1}^{K} \bm H_k \bm A_{k,t} l \left( \sum\limits_{n\in \mathcal{N}_{k,t}} \! \!\!\! \nabla f\left(\bm g_t, \bm x_{n,k}, \bm y_{n,k}\right)\!\right) \!+\!\bm n_t\right)}{\sum\limits_{k = 1}^{K} |\mathcal{N}_{k,t}|} \!\right). 
\end{split}
\end{equation}
This completes the proof.

\section{Latency Models}
\label{app:latency}
We present here the models employed for calculating latency/delay in the experiments. For our algorithm, the latency encompasses the time of updating and transmitting both the MLP and the local FL models. For the analog FL and BPSK FL baselines, the latency consists of the time of updating and uploading the FL parameters only. Our models of the four components are given below.

\textbf{MLP updating time:} The updating time of the MLP depends on the computational complexity of the MLP and the computational power of each device. The computational complexity of the MLP depends on the size of input $\bm g'_{t-1}$ and output ${\rm \Delta}{\widetilde {\bm w}'}_{k,t}$, as well as the number of the neurons in the hidden layer. As discussed in Section III-B, the sizes of $\bm g'_{t-1}$ and ${\rm \Delta}{\widetilde {\bm w}'}_{k,t}$ are both $V'$, and the number of neurons in the hidden layer is $D$. Then, following the latency models in \cite{ZZY}, the updating time of the MLP can be modeled as 
\begin{equation}\label{eq:MPL_inf}
\begin{aligned}
L_{\rm MLP, U}=\frac{2 \alpha^2 V' \! D }{\varepsilon f_{\rm D}} \rho,
\end{aligned}
\end{equation}
where $2V'D$ is the total number of parameters, $f_{\rm D}$ is the CPU clock frequency of a device (in cycles/sec), $\varepsilon$ is the number of parameter update operation executed by one CPU cycle, $\rho$ is the time-consumption coefficient depending on the specific chip of each device, and $\alpha$ is the quantization precision.

\textbf{FL updating time:} The updating time of the FL model similarly depends on the computational complexity of the adopted model and the computational power of each device. For Fashion-MNIST and CIFAR-10, we employ a series of convolutional layers and one linear layer. The training complexity depends on the specific structure of each layer.
%The training complexity of ResNet-18 that consists of seventeen convolutional layers and one linear layer depends on the specific structure of each layer.
In each convolutional layer $i$, let $M_i$ denote the size of the convolutional kernel, $K_i$ denotes the side length of the output feature map generated by each convolutional kernel, and $C^{\rm in}_i$ and $C^{\rm out}_i$ denote the number of input channels and output channels, respectively. Then, the number of operations required for training one of the models is $O_{\rm FL} = \sum\limits_{i = 1}^{I} M^2_i \times K_i^2 \times C^{\rm in}_i \times C^{\rm out}_i + C^{\rm out}_{I+1} \times L_{\rm out}$, where $L_{\rm out}$ is the output dimension of the entire network and $I$ is the number of convolutional layers \cite{xia}. Then, the updating time of the FL model is given by
\begin{equation}\label{eq:FL_train}
\begin{aligned}
L_{\rm FL, U}=\frac{ O_{\rm FL} \alpha^2}{\varepsilon f_{\rm D}} \rho.
\end{aligned}
\end{equation}

\textbf{MLP transmission time:} The transmission time of the MLP depends on the size of the MLP and the channel states. Given the number of parameters is $2 V' D$, the transmission time of the MLP can be modeled as
\begin{equation}\label{eq:MLP_transmit}
\begin{aligned}
L_{\rm MLP, T}=\frac{l\left( 2 \alpha V' D \right)}{C},
\end{aligned}
\end{equation}
where $l\left( \cdot \right)$ denotes the digital pre-processing function described in Sec. II-A. $C$ is the channel capacity (in bits/sec), which is given for MIMO channels as \cite{DTse1}
\begin{equation}\label{eq:MLP_canpa}
\begin{aligned}
C=\mathbb{E}\left[{\rm log }{\;\rm det }\left({\textbf{I}_{N_r}}+\frac{{\rm SNR}}{N_t}\right)\textbf{H}\textbf{H}^*\right],
\end{aligned}
\end{equation}
where ${\rm SNR}=P_0/\sigma^2$ is the common signal-to-noise ratio (SNR) at each receive antenna with $\sigma^2$ being the variance of the additive white Gaussian noise, $N_t$ and $N_r$ are the numbers of transmitting and receiving antennas, respectively, and $\mathbf{H}$ is the MIMO channel vector.

\textbf{FL transmission time:} The transmission time of the FL model depends on the size of the adopted FL model and the channel state. Given the neural network structures, the transmission delay can be modeled as
\begin{equation}\label{eq:FL_transmit}
\begin{aligned}
L_{\rm FL, T}=\frac{l \left( V \alpha \right)}{C},
\end{aligned}
\end{equation}
where $C$ is as defined above and $V = \sum_{i=1}^{I} M_{i}^{2} \times C_{i}^{\mathrm{in}} \times C_{i}^{\text {out }}+C_{I+1}^{\text {out }} \times L_{\text {out }}$ is the number of parameters in the adopted FL model.

For the CIFAR-10 task, $I = 17$, $V = 1.17 \times 10^7$, and $O_{\rm FL} = 1.7 \times 10^{12}$, and for the Fashion MNIST task, $I = 5$, $V = 9.2 \times 10^5$, and $O_{\rm FL} = 1.2 \times 10^{9}$. The value of $\alpha$ is variable for different baselines: for BPSK, $\alpha = 1$ (i.e., single bit precision), for analog FL, $\alpha = 32$ (i.e., full precision computation), and for our proposed method using 64 QAM, $\alpha = 6$. We set $\varepsilon = 10^{5}$, $ \rho = 1$, and $f_{\rm D} = 1$ GHz in our experiments.

%And $f_{\rm D}$ is set to 1 GHz while both $\varepsilon$ and $\rho$ are set to 1.

\bibliographystyle{IEEEbib}
\renewcommand{\baselinestretch}{1}
\bibliography{main}
\end{document}